\documentclass{emulateapj}

\newcommand{\solar}{$_{\odot}$}
\newcommand{\msolar}{M$_{\odot}$}

\newcommand{\mh}{H$_{2}$~}
\newcommand{\mhs}{H$_{2}$}
\newcommand{\um}{$\mu$m~}
\newcommand{\ums}{$\mu$m}

\def\cs33l{[S\kern.2em{\sc iii}] 33.4810 $\mu$m}

\def\h0{{\rm H_0}}

\def\pm{^+_-}
\def\ciil{[C\kern.2em{\sc ii}] 158 $\mu$m }

\def\oii{[O\kern.2em{\sc ii}] }

\def\oiii{[O\kern.2em{\sc iii}] }

\def\oiiil{[O\kern.2em{\sc iii}] 88.36 $\mu$m }

\def\oiiif{[O\kern.2em{\sc iii}] \kern.2em{3p2-3p1} }

\def\oiiie{[O\kern.2em{\sc iii}] \kern.2em{3p1-3p0} }

\def\oiv{[O\kern.2em{\sc iv}] }

\def\siiil{[S\kern.2em{\sc iii}] 18.71 $\mu$m }

\def\sivl{[S\kern.2em{\sc iv}]  10.51  $\mu$m }

\def\s2l{[Si\kern.2em{\sc ii}]  34.8152  $\mu$m }

\def\neiil{[Ne\kern.2em{\sc ii}]  12.81  $\mu$m }
\def\neiiil{[Ne\kern.2em{\sc iii}]  15.56  $\mu$m }

\def\siii{[S\kern.2em{\sc iii}]}

\def\siv{[S\kern.2em{\sc iv}]}

\def\si2{[Si\kern.2em{\sc ii}]}

\def\neii{[Ne\kern.2em{\sc ii}]}
\def\neiii{[Ne\kern.2em{\sc iii}]}
\newcommand{\iraca}{$[3.6] $ }
\newcommand{\iracb}{$[4.5] $ }
\newcommand{\iracc}{$[5.8] $ }
\newcommand{\iracd}{$[8.0] $ }

\shorttitle{Spitzer TDGs}
\shortauthors{Higdon et al.}

\begin{document}

\title{First Detection of PAHs and Warm Molecular Hydrogen in Tidal Dwarf Galaxies}


\author{S. J. U. Higdon \altaffilmark{1}
J. L. Higdon\altaffilmark{1} \& J. Marshall\altaffilmark{1}}

\altaffiltext{1}{Astronomy Department, Cornell University, Ithaca, NY 14853}

\begin{abstract}
  We observed two faint tidal dwarf galaxies (TDGs), NGC~5291~N and
  NGC~5291~S with the Infrared Spectrograph on the Spitzer Space
  Telescope.  We detect strong polycyclic aromatic hydrocarbon (PAH)
  emission at 6.2,~7.7,~8.6,~11.3,~12.6~and~16.5~\ums, which match
  models of groups of $\sim$~100 carbon atoms with an equal mixture of
  neutral and ionized PAHs.  The TDGs have a dominant warm
  $\sim$~140~K dust component in marked contrast to the cooler,
  40~$-$~60~K dust found in starburst galaxies. For the first time we
  detect the low-J rotational lines from molecular hydrogen. Adopting
  LTE there is $\sim$~10$^5$~\msolar~ of $\sim$~400~K gas, which is
  $<$~0.1~\% of the cold gas mass inferred from $^{12}$CO~(1-0)
  measurements. The combination of one-third solar metallicity with a
  recent, $<$~5 million~year, epsiode of star formation is reflected
  in the S and Ne ratios. The excitation is higher than typical values
  for starburst galaxies and similar to that found in BCDs. Using the
  Infared Array Camera we identify an additional 13 PAH-rich candidate
  TDGs. These sources occupy a distinct region of IRAC color space
  with \iraca~$-$~\iracb~$<$~0.4 and \iracb~$-$~\iracd~$>$~3.2. Their
  disturbed morphologies suggest past merger events between
  companions, for example, NGC~5291~S has a projected 11~kpc tail.
  NGC~5291~N and S have stellar masses of
  (1.5~and~3.0)~$\times$~10$^{8}$~M$_{\odot}$, which is comparable to
  BCDs, though still roughly 10$\%$ of the LMC's stellar mass. The
  candidate TDGs are an order of magnitude less massive.  This system
  appears to be a remarkable TDG nursery.

\end{abstract}

\keywords{ galaxies: dwarf --- galaxies: interactions --- individual (
  NGC~5291, NGC~5291~S, NGC~5291~N) --- galaxies: formation ---
  infrared: galaxies }

\section{Introduction}

In addition to triggering starbursts and active galactic nuclei,
mergers of dusty, gas rich disk galaxies frequently lead to the
formation of tidal tails stretching $\ga$100 kpc from the site of the
collision (Toomre \& Toomre 1972; Schweizer 1978;  Sanders \& Mirabel
1996). These structures tend to be HI rich with blue optical colors,
reflecting both their origin in the outer spiral disk and on-going
star formation (van der Hulst 1979; Schombert Wallin \& Struck-Marcell
1990, Mirabel, Lutz \& Mazza 1991; Hibbard \& van Gorkom 1996). Zwicky
(1956) proposed that dwarf galaxies might form out of self-gravitating
clumps within tidal tails, and indeed, concentrations of gas and star
forming regions, with dwarf galaxy size HI masses and optical
luminosities, are commonly found. Computer simulations of merging
galaxies typically give timescales of $\sim 10^8$ years for the
formation of condensations at the tips of the $\sim$ 100 kpc tails,
indicating that any young stellar population has been formed in-situ
and has not simply been stripped from the host galaxy disk.

The cataloging of Tidal Dwarf Galaxies (TDGs) is still in its infancy.
Kinematic observations alleviate potential contamination from
foreground and background objects. However projection effects can give
the appearance of a bound object to unbound condensations in the
debris left over from galaxy interactions. A bona fide TDG should be
self-gravitating, rotating, long-lived and tidal in origin.  Tidal
origin and longevity can only be inferred from comparison of the
source morphology and kinematics with simulations. It is often
impossible to fully disentangle tidal effects from other dynamical
effects such as ram pressure stripping. Whether or not most 
candidate TDGs are subject to tidal disruption and re-absorption by the
merger remnant on $\sim$Gyr timescales is still an open question
(e.g., Hibbard \& Mihos 1995; Elmegreen, Kaufman \& Thomasson 1993).
The formation of both TDGs and transient unbound star-forming regions
will enrich the outer regions of the inter-galaxy medium.

The study of dwarf irregular proto/young galaxies formed either via
tidal interactions between the parent galaxies or from ram-sweeping of
debris material, and collectively called TDGs, may be an important
part of galaxy formation and evolution. TDGs located at the tips of
optically faint but HI rich tails represent new star forming systems
largely free of pre-existing stellar populations, and represent useful
templates of star formation in the early universe, albeit with higher
metallicity (Duc \& Mirabel 1998, hereafter, DM98). More importantly,
dwarf galaxies are by far the most common galaxy type in the current
epoch, accounting for 38 out of 41 galaxies in the Local Group alone
(Mateo 1998).  Work by Hunsberger et al. (1996) suggests that TDGs may
make up at least 1/3 and possibly 1/2 of the dwarf galaxy population
in compact groups i.e., a potentially significant galaxy formation
mechanism in the current epoch.  One problem with this scenario is
dark matter. Like massive spirals, dwarf galaxies appear to be
dominated by dark matter. If dark matter is non-baryonic it is hard to
form a halo around a dwarf galaxy that has been formed via tidal
interaction. If the dark matter is baryonic i.e., cold molecular gas
(Pfenniger, Combes \& Martinet 1994) then the TDG/dwarf association
remains a valid possibility.  Both the nature of dark matter and
whether the dynamical masses of TDGs infers its presence are still
open questions.

Studies of TDG properties may help us better understand the dwarf galaxy
population as a whole, such as what fraction are truly primordial
building blocks of massive galaxies left over from the epoch of galaxy
formation, and what fraction may be TDGs from high and intermediate
red-shifts. This is especially relevant as the merger rate was likely
much higher at z $>$ 1 than it is today.  To date the TDG sample is
still too small to statistically assess the importance of this
population of objects.  That assessment will be made in due course.
Whether rare or ubiquitous, these objects warrant detailed study. Here
we utilize the unique capabilities of the Spitzer Space Telescope
(Werner et al.  2004a) to study the NGC~5291 system, which is known
for its remarkably rich proto/young galaxy nursery.

NGC~5291 is a disturbed elliptical/lenticular galaxy located at the
western edge of the cluster Abell 3574. Pedersen et al. (1978) briefly
discussed deep images of the system taken with the ESO 3.6 m
telescope.  They highlighted the presence of many optical knots, which
they suggested were extragalactic HII regions. Longmore et al. 1979
(hereafter, L79) published a much more detailed investigation.  Using
deep Schmidt plates with filters approximating B ($\Delta\lambda$ 3950
- 5400 \AA~ Schott GG395 filter/III-aJ emulsion) and U
($\Delta\lambda$ 3200 - 3900 \AA~ Schott UG1 filter/III-aJ emulsion)
passbands they found 24 blue knots extending across an arc centered on
NGC~5291 and obtained spectra of the eight brightest. They identified
the knots as giant extragalactic HII regions, but noted that both the
projected linear size and optical luminosity of the largest knots were
greater than the Large Magellanic Cloud. Coincidentally, Maza et al.
(1991) also obtained spectra of the brightest optical knot, which was
included in an object prism survey of HII galaxies. Using the Parkes
radio telescope, L79 also discovered a large HI mass ($\sim$10$^{11}$
M$_{\odot}$) associated with NGC~5291 and its blue knots though
slightly displaced to the west.  Malphrus et al. (1997, hereafter, M97)
mapped the HI 21 cm emission using the VLA.  The HI morphology
resembles a large fragmented ring with a continuous eastern arc
stretching $\sim$ 4 arcmin (72 kpc) on either side of NGC~5291.  The
eastern half of the ring contains most of the neutral hydrogen as well
as the highest HI surface densities, and is associated with both
NGC~5291 and its knots of star formation.  The origin of this giant HI
ring is still unclear.  NGC~5291 has a disturbed companion galaxy
called the ``Seashell'', however it is unlikely that this galaxy is
linked to the HI ring's formation as it is both gas poor and has a
much higher radial velocity ($\sim$600 km s$^{-1}$) relative to
NGC~5291, indicating a high fly-by velocity (L79, $>$ 400 km s$^{-1}$,
DM98). L79 discussed several models for the origin of NGC~5291. They
noted that the optical morphology is similar to the Antennae (NGC
4038/39) and that there are two other galaxies (\# 435 and 446 in
Richter 1984) within 150 kpc projected distance from NGC~5291 that may
be responsible for a tidal interaction.  However M97 rule out the
interplay for both galaxies.  They detected Richter 435, which lies
$\sim$ 4\arcmin~ southwest of NGC~5291, in HI, but with a heliocentric
velocity $\sim$ 250 kms$^{-1}$ lower than the HI ring. M97 found that,
like the Seashell, there is little HI associated with the second
candidate galaxy Richter 446, $\sim$ 5\arcmin~ northeast of NGC~5291.
However, L79's preferred model for the system is that a pre-existing
giant HI disc has undergone compression and possible ram pressure
sweeping by the intra-cluster medium as the system falls into the
cluster from the far side.  L79 suggested that the knots could
eventually form a population of blue dwarf irregular galaxies. The
combined optical and VLA observations led M97 to revive the tidal
model and they conclude that the system is probably both interacting
and likely being ram-swept as it moves through the intra-cluster
medium.

However mysterious its origin the NGC~5291 system is a remarkable
galactic nursery. M97 investigated whether the knots are
self-gravitating and will remain bound against the tidal force of the
parent galaxy to become dwarf irregular galaxies. They compared the HI
mass for each knot to the binding mass. They argued that for the two
most massive HI clumps, NGC~5291~N (hereafter, TDG-N), and NGC~5291~S
(hereafter, TDG-S) the combined mass of stars and gas will result in
self-gravitating knots.  M97 also calculated the radius at which
material in a knot would be disrupted by the tidal force with the
parent galaxy and compared it to the projected radius of the HI knots.
The majority of the knots, including TDG-N and TDG-S, appear to be
tidally stable. Using H$\alpha$ Fabry-Perot observations Borunaud et
al. (2004) were able to determine the rotation curve for TDG-N,
clearly showing that it is self-gravitating and rotating. They
concluded that higher resolution observations are required to assess
the presence of dark matter.

Before the launch of the Spitzer Space Telescope, mid-infrared
spectroscopic studies of objects as faint as TDGs were not feasible.
We have taken advantage of the unprecedented sensitivity of the
Spitzer Space Telescope to obtain IRS spectra of the two optically
brightest TDGs in the NGC~5291 system, TDG-N and TDG-S at the northern
and southern-most tips of the high surface density HI arc. The
mid-infrared region of the spectrum is rich in atomic and molecular
emission features from the star forming ISM, including fine structure
lines like [Ne~II], polycyclic aromatic hydrocarbons (PAHs) and warm
molecular hydrogen.  This will allow comparisons with similar
mid-infrared studies of other star forming galaxies from dwarf
irregulars to massive spirals.

We have also imaged the eastern half of the HI ring, including the
associated proto/young irregular galaxy population, NGC~5291 and
the ``Seashell'', with the Infrared Array Camera (Fazio et al.  2004,
IRAC). We assess the relative contributions
from both cool (evolved and low mass main sequence) stars and from
star formation as traced by PAH emission in the 8.0 $\mu$m band. The
resulting mid-infrared spectral energy distribution (SED) and colors
will allow useful comparisons with other galaxy types.

In $\S$2 we detail the observations and data reduction. $\S$3 begins
with the results and discussion of the the mid-infrared spectral
properties of TDG-N \& TDG-S. We compare the stellar and non-stellar
(i.e., PAH) morphology of NGC~5291, the ``Seashell'', and the knots
next in $\S\S$3.2 $-$ 3.4. In $\S$3.5 we select a population of
candidate TDGs by matching the IRAC SEDs to galaxy templates. We
investigate the mid-infrared color-color and color-magnitude relations
in $\S$3.6. We estimate the TDG stellar masses in $\S$3.7 and present
our conclusions in $\S$4. We adopt a luminosity distance of 62 Mpc to
the NGC~5291 system. This distance is calculated using the red-shift
in NED with H$_{o}$ = 71 kms$^{-1}$Mpc$^{-1}$, $\Omega_M$ = 0.27,
$\Omega_{\Lambda}$ = 0.73. \& $\Omega_{k}$ = 0, and does not reflect
any uncertainty in the distance due to proper motion within the group.
This gives a linear scale of 1\arcsec~= 0.3 kpc.

\section{Observations \& Data Reduction}

\subsection{IRS Spectroscopy}

IRS observations were taken of both TDG-N and TDG-S.
The IRS low resolution data were obtained with Short-low (IRS-SL), which
operates between 5.2 and 7.7 \um (IRS-SL2) and
7.4 - 14.5 $\mu$m (IRS-SL1).  IRS-SL has a resolving
power of 64 $\le$ $\frac{\lambda}{\Delta\lambda}$ $\le$ 128.  The high
resolution data were obtained using Short-high (IRS-SH), which
encompasses the range 9.9 - 19.6 \um, and Long-high (IRS-LH), which
spans 18.7 - 37.2 \ums. The two high resolution spectrometers have a
resolving power, $\frac{\lambda}{\Delta\lambda} \sim 600$.

Observations were made in the IRS Staring Mode AOR with a high
accuracy blue peak-up using a star from the 2MASS catalogue (Cutri
et al. 2003).  The staring mode AOR integrated for half the time listed in
Table 1 at each of the two nominal nod positions on each slit. The
spectral data were processed as far as the un-flat-fielded two
dimensional image using the standard IRS S11 pipeline. See chapter 7
of the Spitzer Observers Manual
(SOM)\footnote{\url{http://ssc.spitzer.caltech.edu/documents/som/}}
for further observing mode and pipeline details.

Both TDG-N and TDG-S have extended emission at 8 \um with respect to
the IRS slits (3.6\arcsec~ IRS-SL, 4.7\arcsec~ IRS-SH and 11.1\arcsec~
IRS-LH). The spectra were extracted and sky subtracted using the
extended source routines in the SMART analysis package (Higdon et al.
2004).  The IRS-SH and IRS-LH date were extracted using the full
aperture and the IRS-SL spectra were extracted using a column of
20\arcsec~ fixed width. The spectra were flat-fielded and
flux-calibrated by extracting and sky subtracting un-flat-fielded
observations of the calibration stars HR 6348 (IRS-SL) and ksi Dra
(IRS-SH, IRS-LH), and dividing these data by the corresponding
template (Cohen 2003) to generate a 1-dimensional relative spectral
response function (RSRF).  The RSRF was then applied to the TARGET
observations to produce the final spectra.  The residual sky was
subtracted from the IRS-SL data using the off-source observations,
which are part of the Staring mode AOR.  A Zodiacal light model from
the Spitzer Planning Observations Tool (SPOT) was scaled to the IRS
slits and subtracted from the IRS-SH and IRS-LH data. An aperture
correction of 0.85 was applied to the IRS-SL spectra as the sources
are extended. In addition, the IRS-LH TDG-N and TDG-S spectra were
scaled by 0.30 and 0.26, respectively in order to stitch to the IRS-SH
spectra. The scale factors are close to the simple geometric aperture
correction for the differing IRS-SH and IRS-LH slits which is 0.215.
However, these sources are not uniformly extended and the two
apertures are orientated $\sim$ 85$^{\circ}$ with respect to each
other.

\subsection{ IRAC Imaging}

IRAC data in all four bands were acquired using a two point map centered
on NGC~5291 ($\alpha_{J2000} = 13^h47^m23.00^s$, $\delta_{J2000} =$
-30$^\circ$25\arcmin30.0\arcsec). The array was aligned to celestial
north and stepped by 180\arcsec.  Three 3 s frames were taken in each
channel using a 12 position Reuleaux dither pattern with a medium
scale factor, resulting in a total integration time of 432 s.  Due to
the footprint of the IRAC arrays the resulting map has a total area of
110.8 arcmin$^2$ with a sub-region of 59.1 arcmin$^2$ sampled by
all 4 channels. In this paper we will only discuss the properties of
sources located in this sub-region. The IRAC data were processed
through version S11.0.2 of the SSC-IRAC pipeline. For more observing
mode and pipeline information see chapter 6 of the SOM.

The IRAC was designed to map both interstellar emission and
starlight. The 3.6 \& 4.5 \um bands primarily trace the stellar mass
distribution, as the emission at these wavelengths is well
approximated by the Rayleigh-Jeans limit of blackbody emission for
both early- and late-type stars.  At 8 \um the mid-IR traces dust
emission, in particular the 7.7 \& 8.6 \um PAH features from the
interstellar medium.  It is straight forward to subtract the stellar
emission from the 8 \um image to produce the PAH/dust image shown in
Figure 1. The method is outlined in Pahre et al. (2004).  The 3.6
and 4.5 \um images are scaled to match the theoretical colors of MO
III stars (\iraca - \iracb $=$ -0.15, \iracb - \iracc $= +0.11$ and
\iracc - \iracd $= +0.04$) at 8 \ums, and subtracted from the 8 \um
image.

SExtractor (Bertin \& Arnouts 1996) was used to generate a catalog of
sources for each IRAC channel. We then selected all sources that were
identified in all four channels with a positional uncertainty of $\le
$ 2 \arcsec. SExtractor was also run on the dust image to find an
additional 14 sources which were not detected in all four channels.
For these sources SExtractor was re-run on the 3.6 through 8.0 \um
images using the source positions determined from the 8 \um image.
Table 4 lists the flux densities in an 8 pixel (9.8 \arcsec) diameter
aperture. A 16 pixel (19.5 \arcsec) diameter aperture was used for a
few sources, including TDG-N and TDG-S. The flux densities were
corrected to a 20 pixel radius aperture using the following scale
factors 1.086 \& 1.017 (3.6 \ums),1.098 \& 1.018 (4.5 \ums),1.097 \&
1.020 (5.8 \ums) and 1.118 \& 1.026 (8.0 \ums) for the 8 and 16 pixel
apertures respectively, as determined by the IRAC team.  The auto-flux
returned from SExtractor is included for objects that are extended
beyond the 19.5 \arcsec~ circular aperture, such as NGC~5291 and the
Seashell. The Vega zero magnitude flux densities in the four IRAC
bands are 277.5, 179.5, 116.6 and 63.1 Jy at 3.6, 4.5, 5.8 and 8.0 \um
respectively.  Stellar sources were removed from the catalog if they
matched one of the following criteria: 2MASS (J - K) $\le $ 0.5
(Eisenhardt 2004), identified as stellar sources in the Hubble Guide
Star Catalogue Version 2.2, or through visual inspection of the IRAC
images.

\section{Results and Discussion}

\subsection{Mid-Infrared Spectral Properties of TDG-N \& TDG-S}

The top of Figures 2 \& 3 show the IRS-SL spectra of TDG-N and TDG-S.
The broad emission features from PAHs at 6.2, 7.7, 8.6 and 11.3 12.6
\um are clearly present. Bright fine structure lines of \sivl and
\neiil emission lines are also apparent in this low resolution
spectrum, as is a weak rotational 0-0 S(3) 9.66 \um line from
molecular hydrogen.  Figure 4 shows the IRS-SL spectrum of TDG-N
overlaid with the IRS-SL spectrum of the reflection nebulae NGC~7023
(Werner et al. 2004b) and the ISO-SWS spectrum of M~82's nucleus (Lutz
et al. 1998), smoothed to the IRS-SL resolution. The overlay spectra
are normalized to TDG-N's flux density at 7 \ums. Here we are using
NGC~7023 to characterize a low ionization star forming region as it is
illuminated by HD 200775, a B2Ve star. M~82 is used to represent the
mid-infrared spectrum of a starburst. The three spectra are remarkably
similar in shape and only a detailed analysis will reveal their
differing environments. To assess whether the PAHs in TDG-N and TDG-S
are similar to those in M~82 we fit the continuum emission in the
IRS-SL spectrum with the thermal radiation from a distribution of
graphite and silicate grains and the PAHs with a series of Drude
profiles. Table 2 lists the resulting PAH fluxes and equivalent
widths.  The relative strength of the 11.3$/$7.7 \um PAHs in TGN-N
(TDG-S) is 0.22 $\pm$ 0.03 (0.23 $\pm$ 0.05) and 6.2$/$7.7 \um PAHs
0.35 $\pm$ 0.06 (0.33 $\pm$ 0.11). These values are consistent with
groups of $\sim$ 100 carbon atoms in an approximately equal mixture of
neutral and ionized PAHs (see Figure 16 in Draine \& Li 2001).  For
comparison, on the same Figure, the PAH ion fraction is close to unity
in M~82 with similar sized clumps of C atoms. In addition to the PAH
features observed in the IRS-SL spectra, we also detect PAH emission
at 13.5 and 16.5 \um in the IRS-SH spectra, the 16.5$/$7.7 \um PAH
ratio is 0.02$\pm$0.01 and 0.10$\pm$0.05 in TDG-N and TDG-S.

Figures 2, bottom \& 3, bottom show the IRS-SH \& IRS-LH spectra. The
fine structure neon and sulfur lines are prominent as well as weak 0-0
S(1) and 0-0 S(2) emission from \mhs.  The spectra flatten out around
$\lambda >$20 \ums.  A similar flattening is seen in the extremely low
metallicity blue compact dwarf galaxy SBS0335$-$052 (Houck et al.
2004b). Using SMART we fit a diluted black-body to the continuum over
the range 15 $-$ 35 \um.  Using a solid angle equal to the IRS-SH
aperture of 1.35 $\times$ 10$^{-10}$ sr gives a temperature T $=$ 139
$\pm$ 6 K. The derived optical depth at 0.55 \um ($\tau_{0.55\mu m}$)
is effectively zero, and $\alpha$ = 1.2 $\pm$ 0.2 where, $\tau =$
$\tau_{0.55\mu m}(0.55/\lambda)^{\alpha}$. A similar result is
obtained for TDG-S with a temperature T $=$ 142 $\pm$ 31 K. A dominant
cooler 40 $-$ 60 K dust component that is present in most starbursts
and spiral galaxies is absent.

Figures 5 \& 6 show the line profiles from the IRS observations. The
line fluxes and equivalent widths for both TDGs are given in Table 3.
No correction has been made for reddening as the optically derived
extinction is low, typically, A$_B$ $\le$ 1 mag.  (L79, Pena, Ruiz \&
Maza 1991, DM98). Geometric effects can also make any extrapolation
from the optical to the infrared highly uncertain. In the following
paragraphs we use the line strengths and ratios to derive physical
properties of the ionized and molecular ISM in the TDGs.

Both the Ne and the S line ratios can be used to constrain the
properties of the underlying starburst. The starburst sample of
Thornley et al. (2000) have typical \neiii$/$\neii~ ratios between
0.05 and 1.0. For example, M 82 has a neon ratio of 0.18. The
exceptions to this range are the two low metallicity dwarf galaxies,
NGC~5253 and II~Zw~40, which both have one-fifth solar metallicity.
These have neon ratios of 3.5 and 12, respectively.  Lowering the
metallicity produces ``hotter'' main sequence stars for a given mass,
and the radiation is harder due to reduced line blanketing and
blocking, hence the neon ratio increases.  The TDG-N and TDG-S have
neon ratios of 2.4 $\pm$0.1 and 1.4 $\pm$ 0.1 respectively, which is
consistent with their one-third solar metallicity (L79, Pena, Ruiz \&
Maza 1991, DM98).  Figure 10 from Thornley et al. (2000) gives an age
for the most recent episode of star formation of $\sim$ 5 million
years for these neon ratios for a galaxy with one-fifth solar
metallicity and log U = $-$ 2.3, where U is the number of ionizing
photons at the surface of the nebula per hydrogen atom. This result is
consistent with the starburst age derived from optical spectra by
DM98. Both TDG-N and TDG-S have slightly higher metallicity, as well
as fewer ionizing photons (i.e., log U $= -3$), as inferred from model
excitation diagrams in Pena, Ruiz \& Maza (1991, for TDG-S we used
emission lines from DM98). Both these effects will result in a younger
burst age using the figure from Thornley et al. (2000).  This is
consistent with the prediction that the stars have formed in-situ and
were not simply dragged from the outer regions of the parent galaxy
during a tidal interaction.

The \siv$/$\siii~ line ratio is another excitation diagnostic.  In
M~82 this ratio is 0.05 (Verma et al. 2003), compared to 0.48 $\pm$
0.02 and 0.23 $\pm$ 0.05 in TDG-N and TDG-S, respectively.  Verma et
al. (2003) plotted the log of the \neiii$/$\neii~ against the log of
the \siv$/$\siii~ line ratio for a sample of twelve starburst and blue
compact dwarf (BCD) galaxies.  Starbursts tend to occupy the bottom
left quadrant of the plot, having low excitation.  The upper right
quadrant is filled with BCDs and galaxies exhibiting Wolf-Rayet
features. The corresponding ratios for TDG-N and TDG-S place them in
the upper right quadrant with moderate excitation. The \siv$/$\siii~
line ratio in TDG-N is close to the value for the overlap region in
the interacting galaxy pair NGC~4038/NGC~4039, but the \neiii$/$\neii~
ratio is a factor two higher (see Figure 4 in Verma et al. 2003).  The
ratios for TDG-S are lower.  The \siv$/$\siii~ line ratio is similar
to the starburst galaxy, NGC~7714 \siv$/$\siii~ = 0.20, but again the
\neiii$/$\neii~ ratio is a factor two higher (Brandl et al.  2004).

The \siii~ 33$/$18.7 \um ratio is a sensitive function of electron
density in the range 300 - 30,000 cm$^{-3}$.  The \siii~ 33 \um line is
detected in TDG-N though with low signal-to-noise. The \siii~ ratio is
0.7 $\pm$ 0.3, which implies an electron density of 600 $\le n_e \le$
3000 cm$^{-3}$ (Figure 1 in Rubin et al.  1994). L79 derived a limit of $<
10^3$ cm$^{-3}$ from their optical spectrum, which is consistent with
our result derived from infrared lines.

We detect weak \mh emission in both TDG-N \& TDG-S. The 0-0 S(2) 12.28
\um and 0-0 S(1) 17.03 \um rotational emission lines have low
signal-to-noise (SNR $\ge$ 3) in the IRS-SH spectra, but the 0-0 S(3)
9.66 \um is also visible in the IRS-SL spectra (see top of Figures 2
\& 3), and all lines, except for the 0-0 (S2) line in TDG-S, are
detected in the individual spectra from both nod positions, adding
weight to the detection of \mh. The average line profiles are shown in
Figures 5 \& 6. To derive the mass of warm molecular hydrogen we
assume that the emission is optically thin. The critical densities of
the low J levels are relatively low (n$_{\rm cr}< 10^3$ cm$^{-3}$) and
we assume that the populations are in LTE.  Adopting an ortho to para
ratio of 3 we show the excitation diagrams in Figures 7.  These are
simply the natural logarithm of the number of molecules divided by the
statistical weight in the upper level of each transition versus the
energy level. For a single component model the data lies on a straight
line and the excitation temperature is the reciprocal of the fit to
the slope. The mass is derived from the S (1) line luminosity and the
excitation temperature. In the following equations we outline the
derivation for the mass of molecular hydrogen. The energy of a given
level, J, is

\begin{equation}
E_J = 85 k J(J+1)
\end{equation}

where k is the Boltzmann constant. The total mass is 

\begin{equation}
M_{Total} = \frac{4}{3} M_o.
\end{equation}

Here M$_o$ is the mass of gas in the ortho state and 

\begin{equation}
M_o = m_{H_2} N_T
\end{equation}

where m$_{H_2}$ is the molecular mass of \mh and N$_T$ is the total number
of molecules.

\begin{equation}
 N_T = \frac{N_J}{f_J}
\end{equation}

N$_J$  is the number of molecules for the J$^{th}$ state

\begin{equation}
N_J = \frac{L(J)}{A_{J}\Delta E_{J}}
\end{equation}

L(J) is the line luminosity, A$_J$ is the Einstein A coefficient,
$\Delta E_{J} =$ h$\nu_J$, where h is Planck's constant and $\nu_J$ is
the frequency of the emission line. f$_J$ is the partition function
for the J$^{th}$ state

\begin{equation}
f_J = \frac{g_J e^{-\frac{E_J}{kT}}}{\sum_{J',ortho} g_{J'} e^{-\frac{E_{J'}}{kT}}}
\end{equation}

where T is the excitation temperature and g$_J$ is the statistical
weight for a given state.

For TDG-N and TDG-S the temperature is 460 $\pm$ 75 K and 400 $\pm$ 40
K. The value is slightly higher than the average value of $\sim$ 350 K
measured by us for a large sample of ULIRGs. Assuming the primordial
helium mass fraction is $\sim$ 0.25 (e.g. Olive \& Skillman 2004), we
derive a total mass of warm gas of (1.0 $\pm$ 0.7) and (1.5 $\pm$ 0.6)
$\times 10^5$ \msolar~ in TDG-N and TDG-S, respectively.  The 0-0 S(0)
limit is consistent with this temperature, but does not rule out a
colder, say 155 K component, which would imply a mass of a few 10$^6$
\msolar.  In TDG-N and TDG-S the warm gas is less than 0.1 \% of the
cold gas mass inferred from $^{12}$CO (1-0) observations (Braine et
al.  2001).  The warm gas fraction is very small compared to the
typically $\sim$ few percent and higher found in starbursts (Rigopolou
et al.  2002).

\subsection{Large Scale Distribution of Non-Stellar Emission }

Figure 8 shows the stunning false color image of NGC~5291. The 3.6 \&
4.5 \um stellar emission is coded blue and green, respectively, and
non-stellar 8\um emission is coded red. TDG-N and TDG-S stand out
prominently in red. There is an excellent correspondence between the
8\um sources and the 24 near-UV bright knots discovered by L79, with
all the sources in their Figure 1, which overlap with our areal
coverage, detected.  Similarly, all of the emission line sources
detected by L79, Maza et al. (1991) and DM98 are visible.  Figure 9
shows the individual IRAC band images.  The non-stellar emission in
these objects almost certainly arises from strong 7.7 $\mu$m and 8.6
$\mu$m PAH features that fall within the IRAC 8.0 $\mu$m filter.  It
is likely that any additional TDG candidates lie within the projected
$\Sigma_{\rm HI}$ = 1 M$_{\odot}$ pc$^{-2}$ contour overlaid on the
dust image (Figure 1).  The IRAC flux densities for the sources
labelled in Figure 1 are listed in Table 4. A substantial fraction
exist in close groups (e.g., \#49-51), while knots \#36,39,41-43 are
arranged along a 30\arcsec~ (9 kpc) arc-like distribution.  The
brightest PAH emitters are clearly associated with regions of higher
$\Sigma_{\rm HI}$ in Figure 1, though, as noted by both L79 and M97,
there is a tendency to avoid the actual peak.  Also note that a clump
of faint knots surrounded by extensive and diffuse non-stellar
emission lies 80\arcsec~ northwest of NGC~5291.  A number of sources
beyond the projected 1 M\solar$/$pc$^2$ HI distribution also exhibit
bright non-stellar emission.  Because of the width of the 8.0 $\mu$m
IRAC filter, knots with bright non-stellar emission but lacking an
optical spectral identification that connects them with the system,
may be at substantially different red-shifts and not associated with
the NGC~5291 system (e.g., DM98 measured red-shifts of 0.07 and 0.2
for sources \#54 \& \#45 in Figure 1).  Further observations will be
required to confirm their physical connection with the NGC~5291
system.

\subsection{Candidate Tidal Dwarf Galaxy Morphology}

In non-stellar emission the knots range in size from essentially
unresolved (D $\la$ 2.5\arcsec~ or $\sim$0.8 kpc) to $\sim$10\arcsec~
(3 kpc) in extent. A number of them show interesting structure.
Sources \#16 and \#17, for example, appear to be connected by a
ribbon-like bridge in Figure 10, while \#47, 49-51 form a close group
of irregularly shaped blobs. L79 noted that faint tails could be
observed extending up to $\sim$20\arcsec~ from several of the brighter
knots.  These are visible in Figure 10 for TDG-N (\#26) \& TDG-S
(\#33).  TDG-S (\#33) possesses the longest tail at $\sim$38\arcsec~
(11 kpc) in projection, with a noticeable brightening in its
southernmost extent which curves off to the East.  Several large arcs
or loops can be seen in the non-stellar images (see Figures 1, 10 \&
11). Two are visible $\sim$20 and 40\arcsec~southeast of NGC~5291's
nucleus and may represent tidal arms emanating from this galaxy. The
expected stellar component may be blended with the diffuse emission.
Additional wispy non-stellar emission is apparent $\sim$ 1.3 ' south
of NGC~5291 that may be a complex tail from the close pair \#47 \& 51.
While the knots invariably have lower flux densities at 3.6 $\mu$m and
4.5 $\mu$m, the more luminous non-stellar sources are also detected in
these bands.  For example, the heads and tails in TDG-S (\#33) and
TDG-N (\#26) are detected in both stellar 3.6 $\mu$m and non-stellar 8
$\mu$m emission, as can be seen in Figure 10.  However, significant
differences can be seen in the stellar and non-stellar structure of
knots like \#47, 49, 50 - 51 and \# 16 \& 17.  The tails and disturbed
morphologies may be signatures of merger events between companion
knots.

\subsection{ NGC~5291 and The Seashell Galaxies}

While originally classified as an elliptical, NGC~5291 was
re-classified as an S0 galaxy by L79 on the basis of weak spiral
structure and a possible dust lane visible in near-ultraviolet light.
Both are consistent with our assessment that the 3.6 and 4.5 \um bands
are dominated by cool stars, as in normal galaxies. Figure 11,
top-left, shows a closeup of NGC~5291 and the Seashell at 3.6 $\mu$m
using a logarithmic stretch, and with contours in Figure 11,
top-right. NGC~5291 has a bright compact stellar core. A slight
twisting of the outer isophotes is apparent, but otherwise the stellar
distribution is quite regular.  The situation is very different in
non-stellar emission, as shown in Figure 11, bottom-left (also with a
logarithmic stretch), where NGC~5291 is dominated by a bright and
marginally resolved nuclear source (R $\sim$ 2\arcsec, peak F$_{\rm
  8.0 \mu m}$ = 225 $\mu$Jy arcsec$^{-2}$) and an asymmetric
20\arcsec~ (6 kpc) diameter ring (F$_{\rm 8.0 \mu m}$ $\sim$ 36
$\mu$Jy arcsec$^{-2}$).  The ring is highly non-uniform, with two
hot-spots north and west of the nucleus and a bifurcation to the
southeast. The intensity of the non-stellar emission is also
noticeably weaker south and west of the nucleus.  Line emission
detected by DM98 at a position 8\arcsec~ north of NGC~5291's nucleus
coincides with this ring. No line emission was detected 4\arcsec~
south of the nucleus where Figure 11, bottom-left, shows the
non-stellar emission to be considerably weaker. From this we deduce
that the ring is actively forming stars, and the non-stellar emission
reflects strong PAH features.  Whether it is the result of a bar
induced resonance (Buta \& Combes 1985) or a collision as in the
Cartwheel ring galaxy (Higdon 1996) can be determined in principle by
detecting expansion using high angular resolution HI observations.

The two faint arcs of non-stellar emission extending 20$''$ south of
NGC~5291's nucleus in the bottom-left panel in Figure 11 resemble tidal
arms seen in other interacting galaxies. There is in addition to these a
counter-spray of dust projecting $\sim$40$''$ (12 kpc) to the northwest
from NGC~5291's nucleus. This can be easily seen in a smoothed version of
the non-stellar emission (5$''$ fwhm kernel) in the bottom-right panel of
Figure 11, where it coincides with $\Sigma_{\rm HI}$$>$ 5 M$_{\odot}$
pc$^{-2}$ gas in Figure 1 and gradually merges into a fainter and more
diffuse non-stellar distribution. Knots of non-stellar emission are also
visible in this material, though they are not as bright at 8.0 $\mu$m as
the majority of the candidate TDGs. The extent of this material suggests a
direct connection between NGC~5291 and the TDGs and candidates even in the
absense of the HI distribution.

The Seashell displays a highly perturbed stellar morphology at 3.6 \um
(Figure 11, top-left), similar to that seen in optical light.  This
consists of a prominent nucleus and edge-on disk plus two structures
extending above the plane, one forming an apparent ring. In
non-stellar emission, only the nucleus and edge-on disk components
remain, with both much fainter than NGC~5291. This is consistent with
its lack of measured HI and line emission (M97; DM98).  Also
noteworthy is the disk galaxy \#5 in Figure 1 (\# 435 Richter 1984,
M97), located $\sim$4\arcmin~southwest of NGC~5291. Its red color in
Figure 8 indicates that it is rich in PAHs with a flocculent spiral
structure.  This galaxy was classified as an SA(s)c by M97.  However a
close inspection of the IRAC images shows evidence of a stellar bar
and a dusty resonance ring. It shows no signs of either a recent tidal
interaction or ram pressure stripping.

\subsection{Mid-infrared Spectral Energy Distribution}

We now have a detailed understanding of the mid-infrared properties of
both TDG-N and TDG-S, which are very similar. Their IRAC SEDs are
``notched'', i,e, the flux density falls from 3.6 to 4.5 \um and then
steeply rises from 4.5 to 8.0 \ums. This spectral shape is
characteristic of star forming regions. For an example, see the
average 2 - 12 \um Infrared Space Observatory PHT-S spectrum from
observations of forty normal galaxies, shown in Figure 3 of Lu et al.
(2003).  As we are searching for additional candidate TDGs, we use the
TDG-N spectrum as our template star forming galaxy.  To characterize
all the sources listed in Table 5, each SED is compared to a series of
template spectra, all normalized at 3.6 \ums.  Including TDG-N the
templates are an S0 (Devriendt, Guiderdoni \& Sadat, R.  1999) an
elliptical (Silva et al.  1998) and the narrow line Seyfert 1, I~Zw~1
from Weedman et al.  (2005) as an AGN template.  The 3.6 - 5 \um
portion of the AGN template is an extension of the continuum to the
IRS 5 - 38 \um spectrum. These templates encompass the observed range
of IRAC SEDs and the closest match is used to classify each source in
Table 4.  For TDG-N and TDG-S we measured the near-infrared J, H and
Ks flux densities using a 19.5\arcsec~ diameter aperture on the 2MASS images
(Cutri et al. 2003). For TDG-S (\# 33) we quote a 5-$\sigma$ upper
limit for the H-band magnitude. To match the extended wavelength
coverage of these two TDGs we included the ISO-SWS spectrum of the
nucleus of the starburst galaxy M~82 (Lutz et al.  1998) with
additional J, H and Ks photometry from the 2MASS images matched to the
ISO-SWS aperture.  TDG-S shows evidence of a 1.6 \um ``bump'' from
evolved stars, whereas the near infrared emission is flatter in TDG-N
and a closer match to M~82, which is shown as a dotted line Figure 12
(dark blue line in electronic edition).  The near-infrared component
is likely from a combined population of evolved and low-mass main
sequence stars.  Figure 12 shows example matches for all our
templates\footnote{SED matches for the full sample are given at
  \url{http://isc.astro.cornell.edu/~sjuh/tdg/NGC5291/}}.  There are
15 objects within the ring (including TDG-N) whose SEDs are well fit
by TDG-N (\# 15, 17, 18, 19, 25, 26$/$TDG-N, 32, 33$/$TDG-S, 36, 39,
41, 42, 47, 49, 50).  These are our candidate TDGs. An additional 7
sources in the ring resemble the TDG-N template, but with weaker PAH
emission (\# 5, 13, 22, 28, 30, 43, 51). These sources are identified
as star forming regions in Table 4. Source \# 5 is the spiral galaxy
SE of NGC~5291 coincident with the HI ring.  Our results show that 61
\% of the knots identified in the HI ring have mid-infrared SEDs
consistent with a young stellar population with 41 \% identified as
candidate TDGs. If the spectral shape of the candidate TDGs match that
of TDG-N and TDG-S both at shorter and longer wavelengths, then the
near-infrared will probe the evolved and low-mass main sequence
stellar population and the mid-infrared will reveal a population of
sources lacking a dominant cool 40 $-$ 60 K dust component.

Three sources outside the ring (\# 45, 54 \& 59) are a good match to
the TDG-N template. These include the two known background starburst
galaxies in DM97 (\# 54 \& \#45). Two sources (\#64, 66) are a reasonable
match to TDG-N. Being outside of the projected HI, these five sources
are listed as star forming regions. Some sources are a poor match to
TDG-N but have ``notched'' SEDs indicating some star formation.
NGC~5291's (\#44) SED matches the S0 template, while the
Seashell's (\#35) SED is consistent with both an elliptical and S0
template.  Source \#20 is an elliptical galaxy. A number of sources,
\#14, 23, 24, 38, 52, 53, 57, 60, 65, 69, may be elliptical galaxies
or foreground stars, and \#2, 9, 10, 21, 61 may be S0 galaxies or
foreground stars. These are labelled E* and S* respectively, in
Table 4.  Outside of the projected HI ring there are a 7 sources that
may be background AGN as their SEDs can be fit with a simple power law
(\#3, 4, 6, 8, 12 56 73).

The number of sources located inside and outside the projected HI ring
are roughly equal, but the ring has only $\sim$ 40 \% the areal
coverage of the background. 44 \% of the sources located within the
ring are good matches to the TDG-N SED compared to 8 \% of the sources
located outside the ring. To estimate the sample contamination we
assume that the three good matches to TDG-N that lie outside the HI
ring are from foreground/background sources. From this number we
infer that approximately two of the candidate TDGs may not be
associated with the NGC~5291 system.

\subsection{Color Magnitude and Color-Color Relations}

In this section we look at the color-color and color-magnitude plots
to see whether TDG-N, TDG-S and the thirteen candidate TDGs occupy a
distinct region of color-color/magnitude space when compared to other
galaxy types. In Figure 13 we plot the \iracb $-$ \iracd versus \iraca
$-$ \iracb color. The vertical and horizontal lines mark the color
zero-points, using the zero-point flux densities listed in $\S\S$ 2.2.
The plot is now divided into four quadrants.  A source with a flat SED
would lie at the crossing point. Sources in the lower left quadrant
have falling SEDs typical of E and S0 galaxies like the Seashell and
NGC~5291 (see \# 35 \& 44 in Figure 12). AGN-like sources, for
example, \#12a shown in Figure 12, with rising SEDs are located in the
upper right quadrant. The lower right quadrant contains sources with
active star formation.  The candidate TDGs, for example, see \#26, 33
\& 42 in Figure 12, are plotted with a framed-diamond symbol. They
have \iracb $-$ \iracd $>$ 3.2 and \iraca $-$ \iracb $<$ 0.4.  Sources
which are a reasonable match to TDG-N, for example see \#13 in Figure
12, but with weaker 8 \um emission and therefore not included in our
candidate TDG list, are plotted as double diamonds. Sources which are
a good match to TDG-N, but lie outside the projected HI ring, are
plotted as solid diamond symbols.  The remaining ``notched'' SED
sources in this quadrant, for example, see \#48a in Figure 12, are a
poor match to TDG-N and are shown as open-diamond symbols.

For comparison the sample of 18 galaxies from Pahre et al. (2004) are
plotted as crosses. This sample contains representative galaxies from
the Hubble morphological sequence and spans E1 through IB(s)m
galaxies. The candidate TDGs do not overlap with this sample. The late
type galaxy NGC 5669 has the reddest \iracb $-$ \iracd color of 2.7 in
the Pahre et al.  (2004) sample, as compared to the average value of
3.8 for the candidate TDGs.  Also included in Figure 13 are the 19 BCD
galaxies from the KPNO International Spectroscopic Survey (KISS),
observed with IRAC by Rosenberg et al. (2005). These sources, shown as
filled circles, were selected from the KPNO International
Spectroscopic Survey (KISS, Salzer et al. 2000) conducted in the NDWFS
Bo\"{o}tes field (Jannuzi \& Dey 1999), and span a wide range in
metallicity (7.8 $\le$ 12 + logO/H $\le$ 9.1) and star formation rate
(0.1 $\le$ SFR $\le$ 2.1 M$_{\odot}$ yr$^{-1}$). Only one BCD (\#2316)
overlaps with the TDG and candidate TDG distribution. This object is
noteworthy for having the highest oxygen abundance and second highest
SFR (1.3 M$_{\odot}$ yr$^{-1}$) among the BCDs. The two next closest
BCDs in Figure 13 (\#2328 \& \#2359 in Rosenberg et al. 2005) have the
highest and third highest SFRs. Otherwise, their properties (i.e.,
H$\alpha$ equivalent widths, metallicity) do not distinguish them from
the remaining BCD sample. The TDGs and candidate TDGs all show
significantly enhanced non-stellar emission, most likely due to PAHs,
relative to normal spirals and even BCD galaxies.

In Figure 14 we plot the \iraca $-$ \iracd color as a function of the
absolute \iracd magnitude. There is no overlap with the Rosenberg et
al. (2004) sample of BCDs. The average for the two TDGs and the 13
candidate TDGs is $\sim$ 1.5 magnitudes redder in \iraca $-$ \iracd
color and $\sim$ 1.9 magnitudes fainter at 8 \ums. The TDGs and
candidate TDGs have a very narrow range in \iraca $-$ \iracd color of
0.7 compared to the BCD sample, which has a spread of 2.5.

\subsection{Stellar Masses For The Candidate TDGs}

A widely used technique for determining the stellar mass distributions
in disk galaxies is to multiply the measured broadband luminosity by
the appropriate mass-to-luminosity ratio (van Albada et
al. 1985). This is generally best carried out at near-infrared
wavelengths, where the effects of dust and massive stars are minimized
(Kranz et al. 2001; Verheijen 2001).  Adopting this approach, we
estimated the stellar masses for TDG-N \& TDG-S by first calculating
their luminosities in the K$_{\rm s}$-band, defined as L$_{\rm K_s}$ =
4$\pi$d$^2$$\nu$F$_{\nu}$ with $\nu$=1.4 $\times$ 10$^{14}$ Hz.
Expressed in terms of the sun's K$_{\rm s}$-band luminosity
($L_{\odot,~K_s}$ = 5.3 $\times$ 10$^{25}$ W), this can be written as
\begin{equation} {\rm L_{\rm K_s} = 3.3 \times 10^{-37} ~d^2 ~F_{\rm
K_s}~~~~(L_{\odot,~K_s}), } 
\end{equation} 
where d is the distance in meters and F$_{\rm K_s}$ is the flux
density in Janskys. For TDG-N \& TDG-S we found L$_{\rm K_s}$ = (3.4
and 6.3) $\times$ 10$^{8}$ L$_{\odot,~K_s}$, respectively.  Stellar
masses were obtained by multiplying these luminosities by K$_{\rm
s}$-band mass-to-luminosity ratios (M$_{*}$/L$_{\rm K_s}$) calculated
by Bell \& de Jong (2001) for model spiral galaxies.  This choice
seemed appropriate given that both the metal abundances (Z $\sim$
1/3 Z$_{\odot}$) and optical colors (B-R $\sim$ 0.6) of the candidate
TDGs in the NGC~5291 system are similar to those found in the disks of
late spiral galaxies (L79; DM98).  Bell \& de Jong found that their
derived M$_{*}$/L depended on the optical color, as was expected.
However, the color dependence is minimized in the near-infrared. For a
B $-$ R = 0.6, their models give M$_*$/L$_{\rm K}$ = 0.3-0.6
M$_{\odot}$/L$_{\odot,~K_s}$.  Using these values results in a stellar
mass of 1.5 $\pm$ 0.6 $\times$ 10$^{8}$ M$_{\odot}$ for TDG-N, which
is consistent with Bournaud et al.'s (2004) estimate obtained from its
blue luminosity and a M/L$_{\rm B}$ of two. For TDG-S we derive a
stellar mass of 3.0 $\pm$ 1.0 $\times$ 10$^{8}$ M$_{\odot}$. These are
roughly 10$\%$ the stellar mass of the LMC (Leroy et al. 2005).

We also estimated stellar masses for the thirteen candidate TDGs in
Table 4 showing evidence of strong PAH emission like TDG-N \& TDG-S.
These are the framed-diamond symbol sources in Figures 13 and 14.
Since none of these sources were detected by 2MASS at J, H, or K$_{\rm
s}$, we calculated F$_{\rm K_s}$ flux densities from their 3.6 $\mu$m
data using the mean (K$_{\rm s}$ $-$ [3.6]) color index for TDG-N \&
TDG-S, which is 0.9 $\pm$ 0.5.  For these sources, F$_{\rm K_s}$ in
the above equation was replaced with F$_{\rm 3.6 \mu m}$ $\times$
10$^{-0.4\times0.9}$ $\times$ (645./277.5), where the last factor
takes into account the different zero points in the two bands. We
derive a mean K$_{\rm s}$ luminosity and stellar mass of 4 $\pm$ 2
$\times$ 10$^{7}$ L$_{\odot,~K_s}$ and 2 $\pm$ 1 $\times$ 10$^{7}$
M$_{\odot}$ respectively for these sources, with the largest source of
uncertainty arising from the observed spread in K$_{\rm s}$ $-$ [3.6]
color. TDG candidate \#32 ($\sim$1\arcmin ~southeast of TDG-N) is the
the most massive of this group with M$_{*}$ = 6 $\pm$ 3 $\times$
10$^{7}$ M$_{\odot}$. Defined this way, TDG-N \& TDG-S possess stellar
masses that are roughly an order of magnitude larger than the average
for the thirteen other candidate TDGs  of the HI ring with [4.5]-[8.0]
$>$ 3.2.

In a similar way, we estimated stellar masses for the nineteen BCD
galaxies from Rosenberg et al. (2005). Four of the BCDs were detected
at K$_{\rm s}$ as well, and their average K$_{\rm s}$ $-$ [3.6] color
(1.0 $\pm$ 0.5) is similar to the mean for TDG-N \& S, with a
comparable spread in value. For these galaxies we either calculated
L$_{\rm \odot, ~K_s}$ and M$_{*}$ directly from their K$_{\rm s}$-band
flux densities, or indirectly using their 3.6 $\mu$m flux densities
and mean K$_{\rm s}$ $-$ [3.6] color. For the latter fifteen objects,
the largest source of uncertainty in the K$_{\rm s}$ $-$ [3.6] colors,
which we take to be the observed spread among the four BCDs detected
at K$_{\rm s}$-band. On average, the BCDs have considerably larger
stellar masses than the TDG-candidate population associated with
NGC~5291, with a median M$_{*}$ = 6 $\pm$ 3 $\times$ 10$^{8}$
M$_{\odot}$.  TDG-N \& TDG-S are close to this average value, however
the other TDG-candidates have stellar masses roughly an order of
magnitude smaller than the typical BCD in Rosenberg et al.'s KISS
sample. These objects are detected to considerably larger distances
than NGC~5291, leading to a possible Malmquist bias. However, we find
a nearly identical median stellar mass for the Rosenberg et al.
sample if we restrict ourselves to those BCDs within 125 Mpc, which is
the distance where our faintest TDG-candidate would have been detected
in all four IRAC bands at 5 $\sigma$. Of the sources in Table 4, only
TDG-N \& TDG-S appear to have stellar masses expected for a typical
BCD.

\section{Conclusions}

We have presented mid-infrared spectra of two tidal dwarf galaxies,
TDG-N \& TDG-S in the NGC~5291 system. Both are PAH rich with features
at 6.2, 7.7, 8.6, 11.3 and 16.5 \ums. The 5 - 14 \um spectra are
remarkably similar in shape to both the starburst galaxy M~82 and the
reflection nebula NGC~7023. However, detailed examination reveals
their differences. The relative PAH strengths are indicative of an
equal mix of ionized and neutral PAHs, with groups of $\sim$ 100
carbon atoms, as opposed to the almost totally ionized PAHs present in
the starburst galaxy M~82. The rise in flux density (F$_\nu$) flattens
off around 20 \um in both TDG-N and TDG-S and the spectra can be fit
with a T $\sim$ 140 K dilute blackbody.  This is in marked contrast
with many spiral and starburst galaxies, such as M~82, whose spectra
peak around 60 - 100 \um due to a dominant cooler dust component with
a temperature $\sim$ 40 K.  In TDG-N the 1.6 \um bump from an evolved
stellar population is absent. Both TDG-N and TDG-S are dominated by
young stellar populations. Their spectra exhibit many fine structure
and molecular lines allowing us to calculate the physical properties
of the star forming regions. The electron density derived from the
\siii~ ratio in TDG-N is 600 $\le n_e \le 3000$ cm$^{-3}$.  Both the
\siv$/$\siii~ and \neiii$/$\neii~ ratios are higher than typical
ratios measured in starburst galaxies and closer to the values
measured for BCD and Wolf-Rayet galaxies. Both metallicity and the age
of a starburst effect the level of excitation.  The TDGs' have
one-third solar metallicity and the \neiii$/$\neii~ line ratios are
consistent with models that have the most recent epoch of star
formation occurring less than five million years ago. This results in
a moderately high excitation. The age of the starburst is consistent
with models which predict that the young stars have formed in-situ and
were not part of the parent galaxy.

For the first time we have the sensitivity to measure the low J
rotational lines from molecular hydrogen and thereby make a direct
measurement of the warm molecular mass. Assuming the gas is in LTE
with an ortho-to-para ratio of 3, we derive temperatures of 460 $\pm$
75 K and 400 $\pm$ 40 K, with a corresponding mass of warm gas of (1.0
$\pm$ 0.7) and (1.5 $\pm$ 0.6) $\times 10^5$ \msolar ~in TDG-N and
TDG-S, respectively. This warm mass is $<$ 0.1 \% of the cold gas mass
inferred from $^{12}$CO (1-0) measurements.

IRAC observations have identified 13 additional sources within the HI
ring that resemble TDG-N and TDG-S. These candidate TDGs occupy a
distinct region of IRAC color space with \iraca $-$ \iracb $<$ 0.4 and
\iracb $-$ \iracd $> $ 3.2 as compared to elliptical, spiral,
irregular and BCD galaxies, which tend to have \iracb $-$ \iracd $<$
3.2. The combined TDG and candidate TDG sample has a very narrow
dispersion in \iraca $-$ \iracd color of $\sim$ 0.7, unlike a sample
of BCDs, which has a spread of $\sim$ 2.5.  Compared to the BCDs the
average candidate TDG is $\sim$ 1.5 magnitudes redder with an \iraca
$-$ \iracd color of 3.6 and $\sim$ 1.9 magnitudes fainter at 8 \ums,
with an absolute magnitude of $\sim$ -21.6.  Approximately 40 \% of
the knots identified in the HI ring fit the TDG template with strong
PAH 7.7 \& 8 \um emission a factor $\sim$ 6 times the luminosity of
the cool stellar population.  Using their 2MASS K$_{\rm s}$-band flux
densities and a starburst M$_{*}/$L$_{\rm K_s}$ appropriate for their
B-R colors, we derive stellar masses of (1.5 $\pm$ 0.6) and (3.0 $\pm$
1.0) $\times$ 10$^{8}$ M$_{\odot}$ for TDG-N and TDG-S respectively.
These values are comparable to similarly defined stellar masses for a
sample of BCDs, though still only $\sim$10$\%$ of the LMC's stellar
mass.  Adopting a mean K$_{\rm s}$ $-$ [3.6] color of 0.9 for the
remaining candidate TDGs, we used their 3.6 $\mu$m luminosities to
derive stellar masses from 0.8-6.0 $\times$ 10$^{7}$ M$_{\odot}$.

The IRAC images of the HI ring show an incredibly rich nursery for
star formation. Many of the candidate TDGs have a disturbed morphology
possibly resulting from previous tidal encounters with fellow
companions.  These are new star forming systems and may be local
analogues of star formation resulting from tidal interactions in the
early universe.

Future work will compare this population of TDGs with our Spitzer
sample of TDG systems, which includes observations of Stephan's
Quintet, Arp 245 and Arp 105.

\acknowledgements

We thank the referee, John Hibbard, for his very careful reading of
our manuscript and good suggestions.  We thank Dr. M. Ashby for many
useful conversations concerning IRAC data analysis. We thank Prof. M.
Haynes for many helpful discussions and encouragement.  We thank Prof.
J. R. Houck for the allocation of valuable IRS guaranteed time to the
TDG program.  This work is based [in part] on observations made with
the Spitzer Space Telescope, which is operated by the Jet Propulsion
Laboratory, California Institute of Technology under NASA contract
1407. Support for this work was provided by NASA through Contract
Number 1257184 issued by JPL/Caltech.  This research has made use of
the excellent NASA/IPAC Extragalactic Database (NED) which is operated
by the Jet Propulsion Laboratory, California Institute of Technology,
under contract with the National Aeronautics and Space Administration;
data products from the Two Micron All Sky Survey, which is a joint
project of the University of Massachusetts and the Infrared Processing
and Analysis Center/California Institute of Technology, funded by the
National Aeronautics and Space Administration and the National Science
Foundation; the SIMBAD database; and archival data from the National
Radio Astronomy Observatory, which is a facility of the National
Science Foundation operated under cooperative agreement by Associated
Universities, Inc.

\references

\reference{} van Albada, T. S., Bachall, J. N., Begeman, K.
              \& Sancisi, R. 1985, \apj, 295, 305

\reference{} Bell, E. F. \& de Jong, R. S. 2001, \apj, 550, 212

\reference{} Bertin, E. \& Arnouts, S. 1996, A\&AS, 117, 393

\reference{} Bournaud, F., Duc, P.-A., Amram, P., Combes, F. \& Gach, J. -L. 2004, A\&A 425, 813

\reference{} Braine, J., Duc, P.-A., Lisenfeld, U., Charmandaris, V. Vallejo, 
O., Leon, S. \& Brinks, E. 2001, A\&A, 378, 51 

\reference{} Brandl et al., 2004 ApJS, 154, 188

\reference{} Buta, R. J. \& Combes, F. 1996, Fundam. Cosmic Physics, 17, 95

\reference{} Cohen, M., Megeath, T.G., Hammersley, P.L., Martin-Luis, F., 
\& Stauffer, J. 2003, \aj, 125, 2645

\reference{} Cutri, R. et al. 2003, 2MASS All-Sky Catalog of Point Sources, VizieR On-line
Data Catalog: II/246. Originally published in: University of
Massachusetts and Infrared Processing and Analysis Center,
(IPAC/California Institute of Technology) (2003)

\reference{} Devriendt, J. E. G., Guiderdoni, B., \& Sadat, R. 1999, \aap, 350, 381

\reference{} Draine, B. T. \& Li, A. 2001, ApJ, 551, 807  

\reference{} Duc, P.-A, \& Mirabel, I. F., 1998 A\&A,  333, 813 (DM98)

\reference{} Eisenhardt, P. R. et al. 2004, ApJS, 154, 48 

\reference{} Elmegreen, B. G., Kaufman, M. \& Thomasson, M. 1993, ApJ, 412, 90

\reference{} Fazio, G. G. et al., 2004 ApJS, 154, 10


 
\reference{} Hibbard, J. E. \& van Gorkom, J. H. 1996, \aj, 111, 655

\reference{} Hibbard, J. E. \& Mihos, J. C. 1995, \aj, 110, 140

\reference{} Higdon, J. L. 1996, \apj, 467, 241

\reference{} Higdon, S. J. U. et al. 2004, PASP, 116, 975 

\reference{} Houck, J. R.  et al., 2004a ApJS, 154, 18

\reference{} Houck, J. R. et al., 2004b ApJS, 154, 211

\reference{} van der Hulst, J. M.  1979, \aap, 71, 131

\reference{} Hunsberger, S. D., Charlton, J., \& Zaritsky, D. 1996, 
\apj, 462, 50

\reference{} Jannuzi, B. \& Dey, A. 1999, ASP Conference Series Vol.~191,
         (R. Weymann, L. Storrie-Lombardi, M. Sawicki, \&
          R. Brunner, eds.), p. 111

\reference{} Kranz, T., Slyz, A. \& Rix, H. 2001, \apj, 562, 164

\reference{} Leroy, A., Bolatto, A. D., Simon, J. D., \& Blitz,
              L. 2005, \apj, 625, 763

\reference{}  Longmore, A. J., Hawarden, T. G., Cannon, R. D., Allen, D. A.,
 Mebold, U., Goss, W. M., \& Reif, K. 1979 MNRAS, 188, 285 (L79)

\reference{} Lu, N. et al. 2003, \apj, 588, 199


 \reference{} Lutz, D., Genzel, R., Kunze, D. ,Spoon, H. W. W., Sturm,
 E., Sternberg, A. \& Moorwood, A. F. M. 1998, In: ASP Conf. Ser. 132,
 Starformation with the Infrared Space Observatory. p.89

\reference{} Malphrus, B. K., Simpson, C. E., Gottesman, S. T. \&
 Hawarden, T. G.  1997, AJ, 114, 1427 (M97)

\reference{} Mateo, M. L.  1998, ARA\&A, 36, 435


\reference{} Maza, J., Ruiz, M. T., Pena, M., Gonzalez, L. \&
              Wischnjewsky, M. 1991, A\&A Suppl., 89, 389

\reference{} Mirabel, I. F., Lutz, D., \& Maza, J. 1991, \aap, 243, 367

\reference{} Olive, K. A. \& Skillman, E. D. 2004, \apj, 617, 29

\reference{} Pahre, M. A., Ashby, M. L. N., Fazio, G. G. \& Willner, 
S. P. 2004, ApJS, 154, 235 
 
\reference{} Pedersen, H., Gammelgaard, P. \& Lausten, S. 1978, The
Messenger, 13, 11

\reference{} Pena, M., Ruiz, M.. T. \& Maza, J. 1991, A\&A, 251, 417

\reference{} Pfenniger, D., Combes, F \& Martinet, L. 1994, A\&A, 285, 79

\reference{} Richter, O. -G. 1984, A\&AS, 58, 131

\reference{} Rosenberg, J. L., Ashby, M. L. N., Salzer, J. J. \&
Huang, J. -S. 2005, \apj, submitted.

\reference{} Rubin, R. H., Simpson, J. P., Lord, S. D.,Colgan, W. J., 
Erickson, E. F. \& Haas, M. R. 1994, ApJ, 420, 772

\reference{} Salzer, J. J. et al. 2000, \aj, 120, 80

\reference{} Sanders, D. B. \&  Mirabel, I. F. 1996, ARAA, 34, 749 

\reference{} Schweizer, F. 1978, IAU Symposium 77, Structure and Properties of
 Nearby Galaxies, p. 279.

\reference{} Silva, L., Granato, G., Bressan, A. \& Danese, L. 1998,
\apj, 509, 103

\reference{} Schombert, J. M., Wallin, J. F. \& Struck-Marcell, C. 1990,
\apj, 99, 497


\reference{} Thornley, M. D., Forster Schreiber, M., Lutz, D. Genzel,
R. Spoon, H.  W. W. \& Kunze, D. 2000, ApJ 539 641

\reference{} Toomre, A. \& Toomre, J. 1972, \apj, 178, 623

\reference{} Verheijen, M. A.  2001, \apj, 563, 694.

\reference{} Verma, A. Lutz, D., Sturm, E., Sternberg, A., Genzel, R.
\& Vacca, W. 2003, A\&A, 403, 829

\reference{} Werner, M. W., et al. 2004a, ApJS, 154, 1

\reference{} Werner, M. W., Uchida, K. I., Sellgren, K., Marengo, M.,
Gordon, K. D., Morris, P. W., Houck, J. R. \& Stansberry, J. A. 2004b,
ApJS, 154, 309

\reference{} Weedman, D. W. et al., 2005, ApJ, Submitted. 

\reference{} Zwicky, F. 1956, Ergebnisse der Exakten Naturwissenchaften,
              29, 344


\clearpage

%
%
%

\begin{deluxetable}{ccccccc} 
\tablecolumns{7} 
\tablewidth{0pc}
\tablecaption{Observational Parameters}
\tablehead{
\colhead{Object}&
\colhead{RA (J2000)}&
\colhead{ Dec (J2000)} & 
\colhead{Date} & 
\colhead{Instrument}&
\colhead{Integration }& 
\colhead{Execution Time} \\
\colhead{}&
\colhead{h:m:s}&
\colhead{d:m:s}&
\colhead{}&
\colhead{}&
\colhead{(s)}&
\colhead{(min)} 
}
\startdata 
NGC5291 & 13:47:23.00 & -30:25:30.00& 2004-02-17&IRAC&3 $\times$ 12&44.87\\ 
TDG-N& 13:47:20.50 & -30:20:51.00&2004-07-17&IRS&&77.81\\
&&&&IRS-SL1&60 $\times$ 4&\\
&&&&IRS-SL2&60 $\times$ 4&\\
&&&&IRS-SH&60 $\times$ 4&\\
&&&&IRS-LH&120 $\times$ 7&\\
TDG-S& 13:47:23.00 & -30:27:30.00&2004-07-17&IRS&&77.73 \\
&&&&IRS-SL1&60 $\times$ 4&\\
&&&&IRS-SL2&60 $\times$ 4&\\
&&&&IRS-SH&60 $\times$ 4&\\
&&&&IRS-LH&120 $\times$ 7&\\
\enddata
\end{deluxetable}

\clearpage



\begin{deluxetable}{cccc} 
\tablecolumns{4} 
\tablewidth{0pc}
\tablecaption{PAH Flux}
\tablehead{
\colhead{  $\lambda_{central}$ }&  
\colhead{Flux}     &            
\colhead{ EW} & 
\colhead{PAH$_x/$PAH${7.7}$ }\\
\colhead{ ($\mu$m)}  & 
\colhead{(10$^{-21}$  Wcm$^{-2}$) } &  
\colhead{($\mu$m) }&
\colhead{}
}
\startdata
TDG-N &&&\\
          6.2 &  13.2 $\pm$ 1.4   &     -5.9  $\pm$   0.7 &    0.35 $\pm$  0.06 \\
          7.7 &      37.2 $\pm$ 5.2 & -12.0 $\pm$ 1.4 & 1.00 \\
          8.6 &  8.1 $\pm$ 1.3   &     -2.1 $\pm$   0.4  &    0.22 $\pm$    0.05 \\
          11.3 &  8.1$\pm$ 0.6     &     -1.7  $\pm$   0.2  &    0.22   $\pm$  0.03 \\
          12.0 &  1.8$\pm$ 0.6   &    -0.4$\pm$  0.2   &   0.05   $\pm$  0.02 \\
          12.6 &  5.7 $\pm$  1.2   &     -1.3  $\pm$     0.3  &    0.15  $\pm$   0.04 \\
          13.5 &  1.5 $\pm$ 0.5  &    -0.4  $\pm$    0.1  &   0.04  $\pm$   0.01 \\
          16.5 \tablenotemark{a} &  0.6 $\pm$ 0.2 & -0.01 $\pm$ 0.01 & 0.02 $\pm$ 0.01 \\
&&&\\
TDG-S &&&\\
           6.2  &   3.3 $\pm$ 1.0    &      6.8 $\pm$      2.1   &     0.33 $\pm$     0.11 \\
          7.7  & 10.0 $\pm$ 1.3   &  12.3 $\pm$ 1.8  & 1.00  \\      
         8.6 &   2.6 $\pm$ 0.7    &      2.5  $\pm$    0.7    &    0.27  $\pm$   0.08 \\
          11.3  &   2.3 $\pm$ 0.1      &     1.54 $\pm$     0.30   &     0.23  $\pm$   0.05 \\
        12.0 &   0.6 $\pm$ 0.2        &   0.4    $\pm$  0.2    &   0.06   $\pm$ 0.03 \\
          12.6  &   1.2 $\pm$ 0.4    &     0.8  $\pm$    0.3   &     0.12  $\pm$   0.04 \\
          13.5  &   0.10 $\pm$ 0.06   &    0.1  $\pm$   0.1   &    0.01  $\pm$  0.01 \\

          16.5 \tablenotemark{a} &  1.0 $\pm$ 0.5 & -0.03 $\pm$ 0.02 & 0.10 $\pm$ 0.05 \\
\enddata

\tablenotetext{a} {From the IRS-SH aperture, uncertainty includes aperture scaling factor between IRS-SL and IRS-SH.}

\end{deluxetable}

%

%
%

\begin{deluxetable}{ccccc} 
\tablecolumns{5} 
\tablewidth{0pc}
\tablecaption{Line Fluxes}
\tablehead{
\colhead{}& 
\colhead{TDG-N}&
 \colhead{}&
\colhead{ TDG-S} &
\colhead{}\\
\colhead{Line}  &
 \colhead{Line Flux} & 
\colhead{EW}& 
\colhead{Line Flux} & 
\colhead{EW}\\
\colhead{} & 
\colhead{(10 $^{-22}$ Wcm$^{-2}$)} & 
\colhead{(\um)}&
\colhead{(10 $^{-22}$ Wcm$^{-2}$)}&
\colhead{(\um)}
}
\startdata
\sivl &  7.40 $\pm$  0.15 & -0.018$\pm$ 0.002& 1.05 $\pm$  0.22& -0.002$\pm$ 0.001\\
\neiil & 10.65 $\pm$  0.06 & -0.024$\pm$ 0.001& 4.05 $\pm$  0.31& -0.011$\pm$ 0.003\\
\neiiil & 25.31 $\pm$  1.03 & -0.065$\pm$ 0.002& 5.47 $\pm$  0.11& -0.016$\pm$ 0.002\\
\siiil & 15.45 $\pm$  0.58 & -0.041$\pm$ 0.002& 4.66 $\pm$  0.17& -0.015$\pm$ 0.000\\
\siii & 10.61 $\pm$  4.02 & -0.100$\pm$ 0.049&$\le$37.87 $\pm$  0.54& \nodata \\
0-0 S(3) 9.66\tablenotemark{a} & 2.2  $\pm$ 0.9   & -0.04 $\pm$ 0.02 &  1.6$\pm$0.5 \tablenotemark{b} & -0.08 $\pm$  0.03\\
0-0 S(2) 12.28 &  1.9 $\pm$  0.8  & -0.003$\pm$ 0.001& 0.9 $\pm$ 0.4\tablenotemark{c} & -0.002$\pm$0.001\\
0-0 S(1) 17.03 &  1.1 $\pm$ 0.4  & -0.003$\pm$0.001& 1.3 $\pm$  0.3\tablenotemark{d} & -0.002$\pm$ 0.001\\
0-0 S(0) 28.22 & $\le$ 1.7\tablenotemark{d} & \nodata& $\le$1.7\tablenotemark{e} & \nodata \\
\enddata
\tablenotetext{a} {IRS-SL} 
\tablenotetext{b} {FWHM fixed to instrumental resolution of 0.088 \ums.}  
\tablenotetext{c} {FWHM fixed to instrumental resolution of 0.020 \ums.}  
\tablenotetext{d} {FWHM fixed to instrumental resolution of 0.028 \ums.}  
\tablenotetext{e} {3-$\sigma$ upper limit. Calculated using a 0-order fit to the continuum. The upper limit is 3 $\times$ RMS $\times$ FWHM. The FWHM is 0.05 \ums.}

\end{deluxetable}
\clearpage





\begin{deluxetable}{clcccrrrr}
\tablecolumns{9} 
\tablewidth{0pc}
\tablecaption{IRAC Flux Densities in an 8 Pixel Diameter Aperture}
\tablehead{
\multicolumn{3}{c}{Object\tablenotemark{a}}&
\colhead{RA (J2000)}&
\colhead{Dec (J2000)}& 
\colhead{F$_{3.6 \mu m}$}&
\colhead{F$_{4.5 \mu m}$}&
\colhead{F$_{5.8 \mu m}$}&
\colhead{F$_{8.0 \mu m}$}\\
\colhead{}&\colhead{}&\colhead{}&
\colhead{($^\circ$)}&
\colhead{($^\circ$)}&
\colhead{($\mu$Jy)}&
\colhead{($\mu$Jy)}&
\colhead{($\mu$Jy)}
&\colhead{($\mu$Jy)}
}
\startdata 
                                       1&&AGN&206.77913&-30.463076&82.6$\pm$3.5&67.6$\pm$3.4&73.2$\pm$15.2&67.1$\pm$10.3\\
                   2\tablenotemark{b}&&S0*&206.78200&-30.411720&220.0$\pm$3.5&201.0$\pm$3.4&104.0$\pm$15.2&107.0$\pm$10.3\\
                  3\tablenotemark{b}&&AGN&206.78366&-30.425823&114.0$\pm$3.5&123.0$\pm$3.4&167.0$\pm$15.2&147.0$\pm$10.2\\
                 4\tablenotemark{b}&&AGN&206.80237&-30.392924&126.0$\pm$2.6&175.0$\pm$2.9&268.0$\pm$10.9&365.0$\pm$9.2\\
           5\tablenotemark{cd}&&SF&206.80253&-30.456886&5460.0$\pm$20.2&3390.0$\pm$18.1&6200.0$\pm$65.4&16500.0$\pm$55.3\\
                 6\tablenotemark{b}&&AGN&206.80267&-30.439257&188.0$\pm$3.5&343.0$\pm$3.4&614.0$\pm$15.2&1240.0$\pm$10.3\\
                       7\tablenotemark{b}&&N&206.80351&-30.363409&52.0$\pm$3.5&33.2$\pm$3.4&70.1$\pm$14.9&62.3$\pm$10.3\\
                     8\tablenotemark{b}&&AGN&206.80364&-30.372922&66.8$\pm$3.5&97.1$\pm$3.4&206.0$\pm$15.2&240.0$\pm$10.3\\
                                      9&&S0*&206.80908&-30.468743&107.0$\pm$3.5&74.8$\pm$3.4&79.5$\pm$15.2&61.6$\pm$10.3\\
                   10\tablenotemark{b}&&S0*&206.81047&-30.420336&169.0$\pm$3.5&122.0$\pm$3.4&90.8$\pm$15.2&77.0$\pm$10.3\\
                                   11&&AGN?&206.82215&-30.339189&112.0$\pm$3.5&112.0$\pm$3.4&301.0$\pm$14.3&161.0$\pm$10.0\\
                   12\tablenotemark{b}&&AGN&206.82216&-30.414722&35.2$\pm$2.6&44.0$\pm$2.9&60.2$\pm$10.9&100.0$\pm$9.2\\
                                  13&H -&SF&206.82294&-30.472132&140.0$\pm$3.5&107.0$\pm$3.4&170.0$\pm$15.2&419.0$\pm$10.2\\
                                    14&&E*&206.82355&-30.453369&204.0$\pm$3.5&127.0$\pm$3.4&71.6$\pm$15.2&51.0$\pm$10.2\\
                                     15&- f&TDG&206.82406&-30.366257&39.9$\pm$3.5&29.1$\pm$3.4&86.6$\pm$15.2&252.0$\pm$10.2\\
                                    16&&N&206.82418&-30.356122&145.0$\pm$3.5&122.0$\pm$3.4&92.3$\pm$15.2&231.0$\pm$10.2\\
                                     17&- h&TDG&206.82499&-30.352538&89.9$\pm$3.5&57.1$\pm$3.4&259.0$\pm$15.2&663.0$\pm$10.2\\
                                    18&&TDG&206.82504&-30.364186&45.0$\pm$3.5&24.8$\pm$3.4&130.0$\pm$15.2&352.0$\pm$10.3\\
                                      19&&TDG&206.82672&-30.344119&30.7$\pm$3.5&9.6$\pm$3.4&$<$ 36.3&188.0$\pm$10.2\\
                20\tablenotemark{b}&1 C&E&206.82682&-30.383323&1050.0$\pm$6.5&625.0$\pm$6.2&341.0$\pm$28.0&246.0$\pm$18.5\\
                                    21&&S0*&206.82883&-30.452879&207.0$\pm$3.5&133.0$\pm$3.4&93.7$\pm$15.2&75.0$\pm$10.2\\
                                    22&- b&SF&206.83111&-30.448559&49.5$\pm$2.6&32.1$\pm$2.9&40.7$\pm$10.8&106.0$\pm$9.2\\
                                   23&&E*&206.83137&-30.415181&243.0$\pm$3.5&167.0$\pm$3.4&118.0$\pm$15.2&73.0$\pm$10.3\\
                                  24&&E*&206.83159&-30.399068&471.0$\pm$3.5&316.0$\pm$3.4&224.0$\pm$15.2&101.0$\pm$10.2\\
                                   25&A j&TDG&206.83236&-30.339936&69.3$\pm$2.6&38.0$\pm$2.9&402.0$\pm$10.8&486.0$\pm$9.2\\

               26\tablenotemark{de}&B i&TDG&206.83508&-30.347620&513.0$\pm$6.5&389.3$\pm$6.2&1844.5$\pm$28.2&4804.0$\pm$18.8\\
                                     27&&N&206.83516&-30.447661&80.7$\pm$2.6&56.6$\pm$2.9&28.9$\pm$10.8&76.4$\pm$9.2\\
                                      28&&SF&206.83619&-30.445187&19.8$\pm$2.6&8.1$\pm$2.9&18.9$\pm$10.9&71.1$\pm$9.3\\
                                     29&&&206.83759&-30.384396&$<$11.7&$<$9.0&26.9$\pm$10.8&85.2$\pm$9.2\\
                                 30&4 -&SF&206.83873&-30.472597&373.0$\pm$3.5&256.0$\pm$3.4&286.0$\pm$15.2&1350.0$\pm$10.3\\
                                     31&&&206.84210&-30.383212&8.9$\pm$2.6&10.4$\pm$2.9&52.3$\pm$10.9&158.0$\pm$9.2\\
                                   32&- g&TDG&206.84312&-30.362525&118.0$\pm$3.5&72.5$\pm$3.4&300.0$\pm$15.2&853.0$\pm$10.2\\
                33\tablenotemark{df}&F a&TDG&206.84521&-30.458553&295.0$\pm$6.5&160.9$\pm$6.1&617.8$\pm$28.2&1930.0$\pm$18.8\\

                    34\tablenotemark{b}&&AGN?&206.84561&-30.491370&92.7$\pm$3.5&80.8$\pm$3.4&78.2$\pm$15.2&111.0$\pm$10.3\\
         35\tablenotemark{gh}&- B&E S0&206.84707&-30.417140&17500.0$\pm$21.2&10400.0$\pm$19.9&8050.0$\pm$72.0&6110.0$\pm$54.7\\
                                    36&&TDG&206.84741&-30.449865&27.4$\pm$2.6&17.8$\pm$2.9&28.4$\pm$10.9&174.0$\pm$9.2\\
                    37\tablenotemark{b}&&AGN?&206.84765&-30.489807&61.1$\pm$2.6&59.6$\pm$2.9&48.7$\pm$10.8&72.1$\pm$9.2\\
                                  38&&E*&206.84767&-30.357765&895.0$\pm$3.5&578.0$\pm$3.4&432.0$\pm$15.2&227.0$\pm$10.3\\
                                    39&&TDG&206.84773&-30.447670&17.1$\pm$2.6&15.0$\pm$2.9&27.3$\pm$10.8&113.0$\pm$9.3\\
                    40\tablenotemark{b}&&AGN?&206.84823&-30.345555&51.4$\pm$3.5&79.7$\pm$3.4&52.3$\pm$15.2&101.0$\pm$10.2\\
                                    41&- c&TDG&206.84868&-30.445188&25.0$\pm$2.6&10.8$\pm$2.9&40.0$\pm$10.9&137.0$\pm$9.2\\
                                    42&E -&TDG&206.84913&-30.452056&51.0$\pm$3.5&32.7$\pm$3.4&105.0$\pm$15.2&258.0$\pm$10.2\\
                                    43&D d&SF&206.84915&-30.439878&62.8$\pm$2.6&40.5$\pm$2.9&62.1$\pm$10.9&180.0$\pm$9.2\\
       44\tablenotemark{gi}&- A&E&206.85203&-30.407169&69400.0$\pm$41.9&40800.0$\pm$36.1&30900.0$\pm$87.2&36400.0$\pm$88.7\\
                   45\tablenotemark{b}&- 2&SF&206.85413&-30.374597&76.0$\pm$2.6&62.0$\pm$2.9&25.1$\pm$10.8&198.0$\pm$9.2\\
                   46\tablenotemark{b}&&&206.85697&-30.492522&282.0$\pm$3.5&176.0$\pm$3.4&71.9$\pm$15.2&60.9$\pm$10.3\\
                                    47&&TDG&206.85752&-30.430211&38.6$\pm$2.6&17.3$\pm$2.9&97.6$\pm$10.8&284.0$\pm$9.2\\
                48\tablenotemark{b}&&N&206.85858&-30.395310&205.0$\pm$2.6&181.0$\pm$2.9&149.0$\pm$10.9&290.0$\pm$9.2\\
                                    49&&TDG&206.85934&-30.433200&27.5$\pm$2.6&15.1$\pm$2.9&50.1$\pm$10.8&158.0$\pm$9.2\\
                                    50&&TDG&206.86013&-30.434832&27.7$\pm$2.6&21.0$\pm$2.9&57.2$\pm$10.8&162.0$\pm$9.3\\
                                  51&C e&SF&206.86030&-30.430790&134.0$\pm$2.6&88.9$\pm$2.9&142.0$\pm$10.8&313.0$\pm$9.2\\
                  52\tablenotemark{b}&&E*&206.86159&-30.508598&215.0$\pm$3.5&175.0$\pm$3.4&121.0$\pm$15.2&76.7$\pm$10.3\\
                  53\tablenotemark{b}&&E*&206.86201&-30.359309&268.0$\pm$3.5&173.0$\pm$3.4&104.0$\pm$15.2&75.2$\pm$10.3\\
                54\tablenotemark{b}&2 1&54&206.86375&-30.371721&663.0$\pm$3.5&446.0$\pm$3.4&629.0$\pm$15.2&4070.0$\pm$10.3\\
                  55\tablenotemark{b}&&&206.86543&-30.447166&115.0$\pm$3.5&111.0$\pm$3.4&104.0$\pm$15.2&64.9$\pm$10.2\\
                   56\tablenotemark{b}&&AGN&206.86557&-30.418478&78.6$\pm$2.6&88.4$\pm$2.9&83.4$\pm$10.9&208.0$\pm$9.2\\
                   57\tablenotemark{b}&&E*&206.86574&-30.367851&173.0$\pm$3.5&116.0$\pm$3.4&73.0$\pm$15.2&58.6$\pm$10.2\\
                   58\tablenotemark{b}&&AGN?&206.86612&-30.414799&14.6$\pm$2.6&19.4$\pm$2.9&17.8$\pm$10.8&150.0$\pm$9.2\\
                 59\tablenotemark{b}&&SF&206.87124&-30.476743&236.0$\pm$3.5&156.0$\pm$3.4&252.0$\pm$15.2&944.0$\pm$10.2\\

\enddata

\end{deluxetable}

\clearpage

\begin{deluxetable}{clcccrrrr}
\tablenum{4}
\tablecolumns{9} 
\tablewidth{0pc}
\tablecaption{- cont.}
\tablehead{
\multicolumn{3}{c}{Object\tablenotemark{a}}&
\colhead{RA (J2000)}&
\colhead{Dec (J2000)}& 
\colhead{F$_{3.6 \mu m}$}&
\colhead{F$_{4.5 \mu m}$}&
\colhead{F$_{5.8 \mu m}$}&
\colhead{F$_{8.0 \mu m}$}\\
\colhead{}&\colhead{}&\colhead{}&
\colhead{($^\circ$)}&
\colhead{($^\circ$)}&
\colhead{($\mu$Jy)}&
\colhead{($\mu$Jy)}&
\colhead{($\mu$Jy)}
&\colhead{($\mu$Jy)}
}
\startdata

                 60\tablenotemark{b}&3 -&E*&206.87193&-30.404070&809.0$\pm$3.5&480.0$\pm$3.4&293.0$\pm$15.2&222.0$\pm$10.3\\
                 61\tablenotemark{b}&&S0*&206.87499&-30.370020&407.0$\pm$3.5&306.0$\pm$3.4&258.0$\pm$15.2&221.0$\pm$10.3\\
                   62\tablenotemark{b}&&N&206.87787&-30.404201&95.5$\pm$2.6&86.7$\pm$2.9&42.1$\pm$10.9&124.0$\pm$9.2\\
                    63\tablenotemark{b}&&N&206.87844&-30.455269&80.0$\pm$2.6&68.9$\pm$2.9&$<$16.8&162.0$\pm$9.3\\
                  64\tablenotemark{b}&&SF&206.88901&-30.390435&133.0$\pm$3.5&117.0$\pm$3.4&96.0$\pm$15.2&284.0$\pm$10.3\\
                 65\tablenotemark{b}&&E*&206.89271&-30.364745&923.0$\pm$3.5&650.0$\pm$3.4&404.0$\pm$15.2&261.0$\pm$10.3\\
                    66\tablenotemark{b}&&SF&206.89412&-30.467691&69.3$\pm$3.2&64.5$\pm$3.3&31.1$\pm$14.9&135.0$\pm$10.2\\
                   67\tablenotemark{b}&&&206.89467&-30.431946&119.0$\pm$3.3&82.9$\pm$3.2&87.9$\pm$15.2&111.0$\pm$10.2\\
                    68\tablenotemark{b}&&N&206.89631&-30.437985&60.2$\pm$2.6&53.1$\pm$2.9&28.5$\pm$10.9&87.7$\pm$8.8\\
                  69\tablenotemark{b}&&E*&206.89914&-30.395302&201.0$\pm$3.5&133.0$\pm$3.3&102.0$\pm$15.2&54.6$\pm$10.3\\
                 70\tablenotemark{b}&&&206.90090&-30.367889&241.0$\pm$3.5&158.0$\pm$3.3&124.0$\pm$15.2&175.0$\pm$10.2\\
                  71\tablenotemark{b}&&N&206.90105&-30.460684&243.0$\pm$3.5&151.0$\pm$3.4&75.6$\pm$15.2&131.0$\pm$10.2\\
                    72\tablenotemark{b}&&&206.90456&-30.421585&37.8$\pm$3.5&49.9$\pm$3.3&36.5$\pm$14.4&114.0$\pm$10.2\\
                     73\tablenotemark{b}&&AGN&206.91035&-30.448771&45.7$\pm$3.5&43.5$\pm$3.4&82.6$\pm$15.2&94.7$\pm$10.2\\
                     74\tablenotemark{b}&&&206.91079&-30.450288&22.7$\pm$3.2&65.0$\pm$3.3&$<$41.2&73.5$\pm$9.7\\
                  75\tablenotemark{b}&&N&206.91286&-30.439546&142.0$\pm$3.5&129.0$\pm$3.4&36.4$\pm$15.2&204.0$\pm$10.0\\
\enddata
\tablenotetext{a} {Second column is the L79/M97
 and DM98 naming convention. Third column is the classification from the SED match in \SS 3.5, T: candidate TDG, 
SF: reasonable fit to TDG-N, star-forming region, N:``notched'' SED but poor match to TDG-N, 
E: elliptical, E*: elliptical/foreground star, S0 : S0, S0* : S0/foreground star, AGN :AGN, 
AGN? : not a simple power-law} 
\tablenotetext{b} {Source is outside HI ring.}
\tablenotetext{c} {Galaxy 435 (Richter 1984, M97).}
\tablenotetext{d} {16 pixel diameter aperture.}
\tablenotetext{e} {TDG-N. Source K09.09 (Pena, Ruiz \& Maza 1991)}
\tablenotetext{f} {TDG-S., the tail is source G in L79}
\tablenotetext{g} {Mag. auto flux density as source is extended beyond 16 pixel diam. aperture. Note the aperture correction for an infinitely extended source has not been applied. The correction factors listed in the SOM are 0.94, 0.94, 0.63, 0.69 at \iraca, \iracb, \iracc, \iracd, respectively.}
\tablenotetext{h} {Seashell.}
\tablenotetext{i} {NGC 5291.}

\end{deluxetable}

\clearpage



\begin{figure}
\centering
\epsscale{2.0}
\caption{The IRAC ``non-stellar'' image of the NGC~5291 system created
  by subtracting stellar light from the 8.0 $\mu$m~image (see text).
  This emission is most likely dominated by strong 7.7 $\mu$m and 8.6
  $\mu$m PAH bands. Arrows identify sources listed in Table 4. For
  example, \# 26 \& 33 are TDG-N and TDG-S, respectively. NGC~5291 is
  \# 44 and the Seashell is \#35. The contours show the distribution of HI
  ($\Sigma_{\rm HI}$ = 1, 3, 5, 7 \& 9 M$_{\odot}$ pc$^{-2}$) created
  by reprocessing the VLA DnC archive data ($\theta_{\rm FWHM}$ =
  50\arcsec). Original VLA data published by Malphrus et al. (1997). }
\end{figure}

\clearpage


\begin{figure}
\centering
\includegraphics[angle=0,scale=.75]{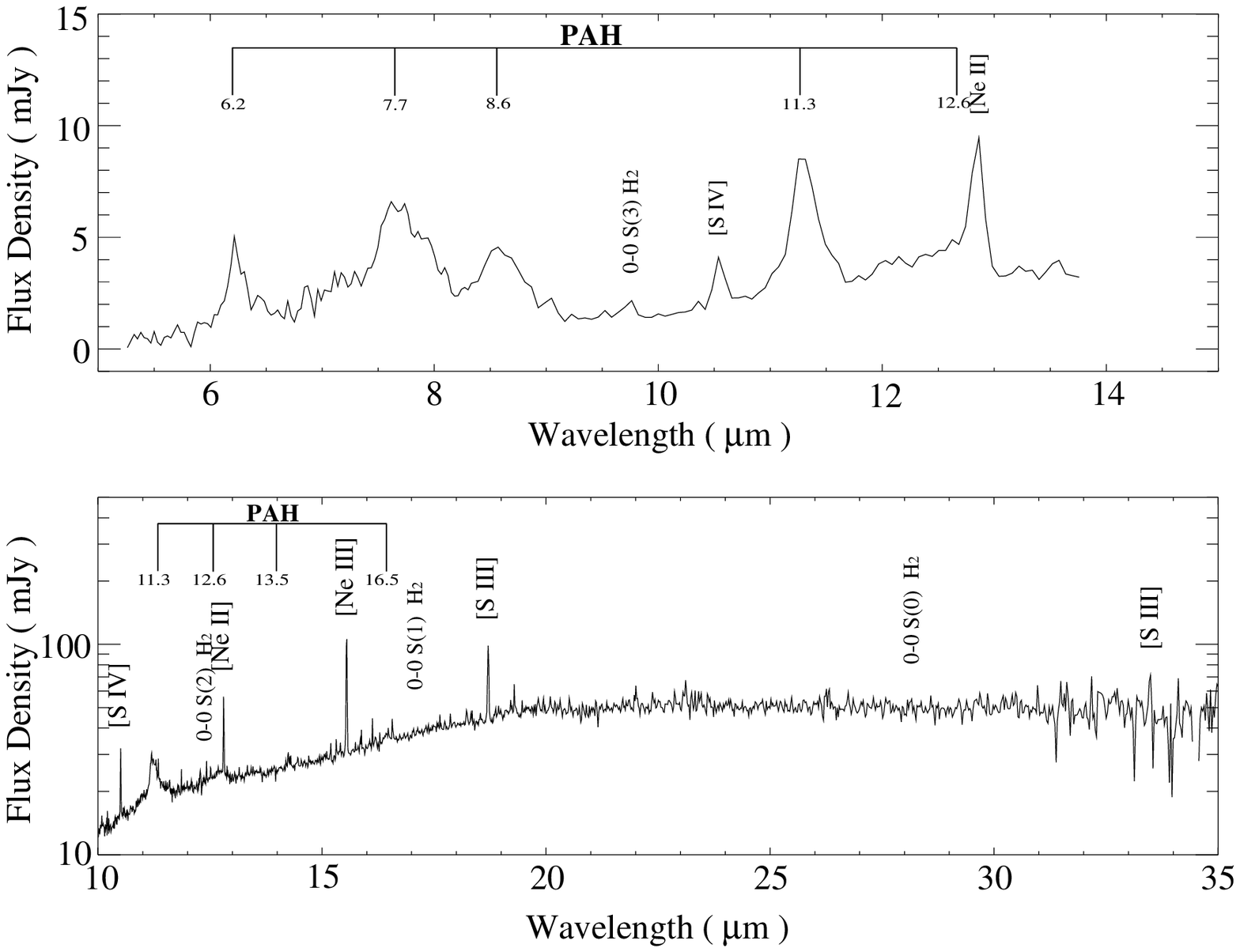}
\caption{Spitzer IRS spectra of TDG-N. (Top) IRS-SL spectrum showing
  strong PAH features and 0-0 S(3) 9.7 \um molecular Hydrogen.
  (Bottom). Combined IRS-SH \& IRS-LH spectrum showing PAHs and fine
  structure lines. The flattening of the continuum emission beyond
  $\lambda >$ 20 \um signifies the absence of a dominant cool 40 K
  dust component. }
\end{figure}

\clearpage


\begin{figure}
\centering
 \includegraphics[angle=0,scale=.75]{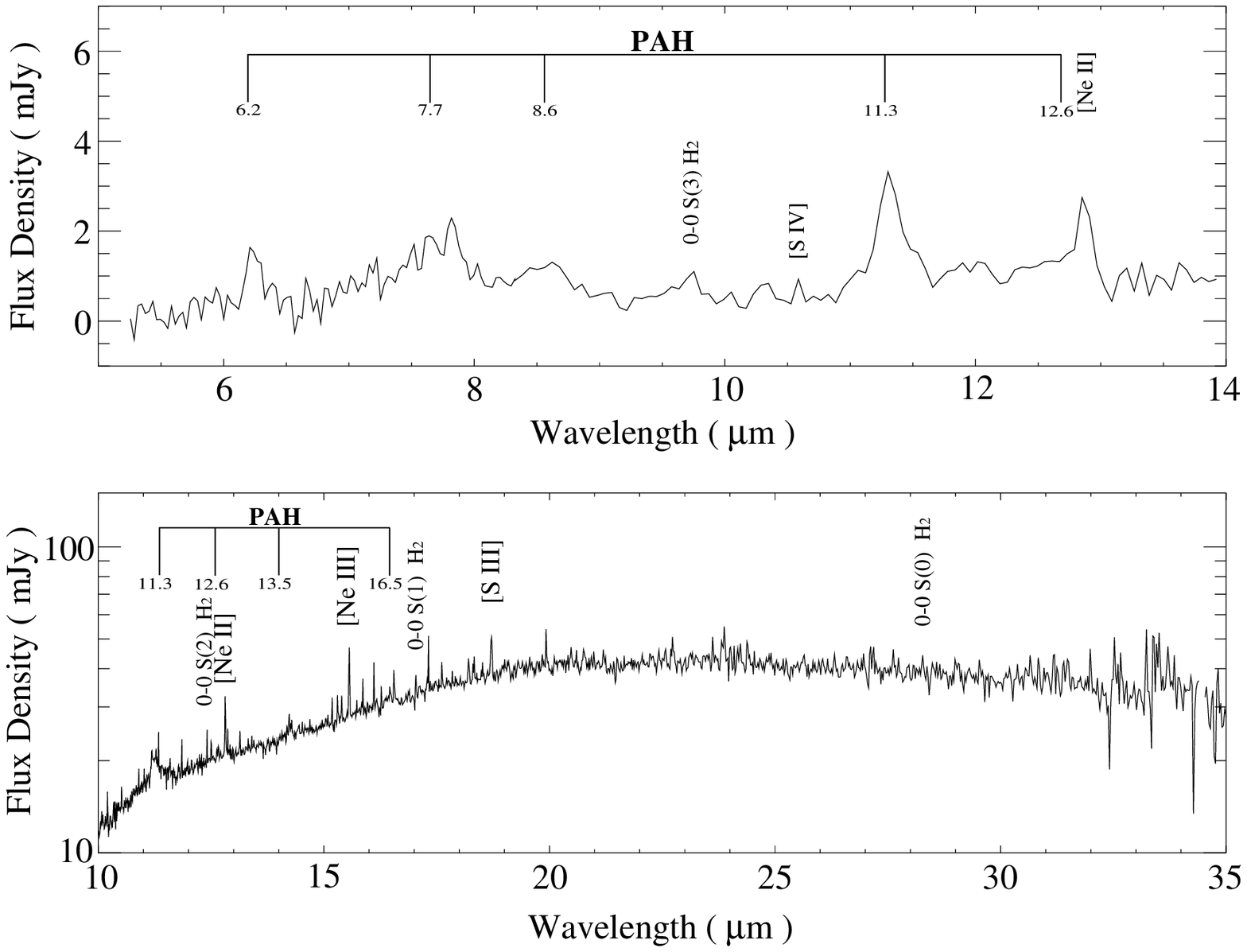}
 \caption{Spitzer IRS spectra of TDG-S. (Top) IRS-SL spectrum showing
   strong PAH features and 0-0 S(3) 9.7 \um molecular Hydrogen.
   (Bottom). Combined IRS-SH \& IRS-LH spectrum showing PAHs and fine
   structure lines. The flattening of the continuum emission beyond
   $\lambda >$ 20 \um signifies the absence of a dominant cool 40 K
   dust component. }
\end{figure}

\clearpage

\begin{figure}
\centering
 \includegraphics[angle=0,scale=.75]{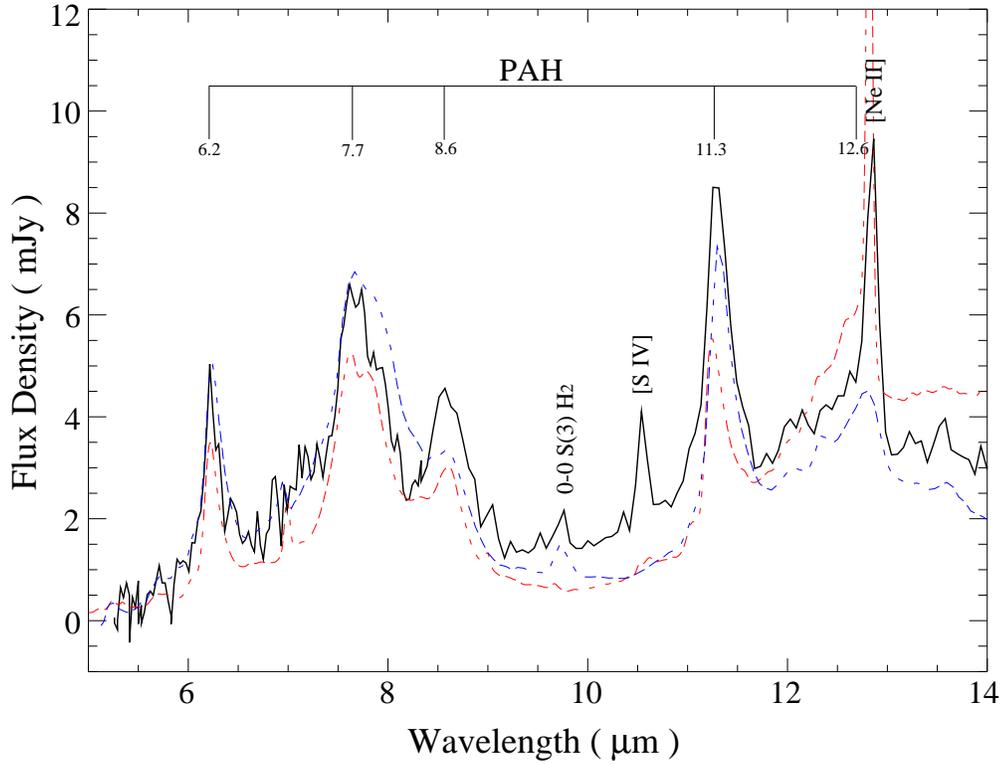}
 \caption{ A comparison between the IRS Short-Low spectrum of TDG-N
   (thick black line), the ISO-SWS spectrum of the starburst galaxy
   M~82 dash line (dashed red-line electronic edition) from Lutz et
   al.  (1998) after smoothing to the IRS Short-Low resolution and
   normalized to 7 \ums, and an IRS Short-Low spectrum of the
   reflection nebula NGC~7023 normalized to 7 \um thin line (dashed
   blue-line electronic edition) from Werner et al.  2004b, showing the
   remarkable similarity in shape.  Note the \neiil line is blended
   with the 12.6 \um PAH. }
\end{figure}

\clearpage


\begin{figure}
\centering
\label{fig:graphics:b}
 \includegraphics[angle=90,scale=.75]{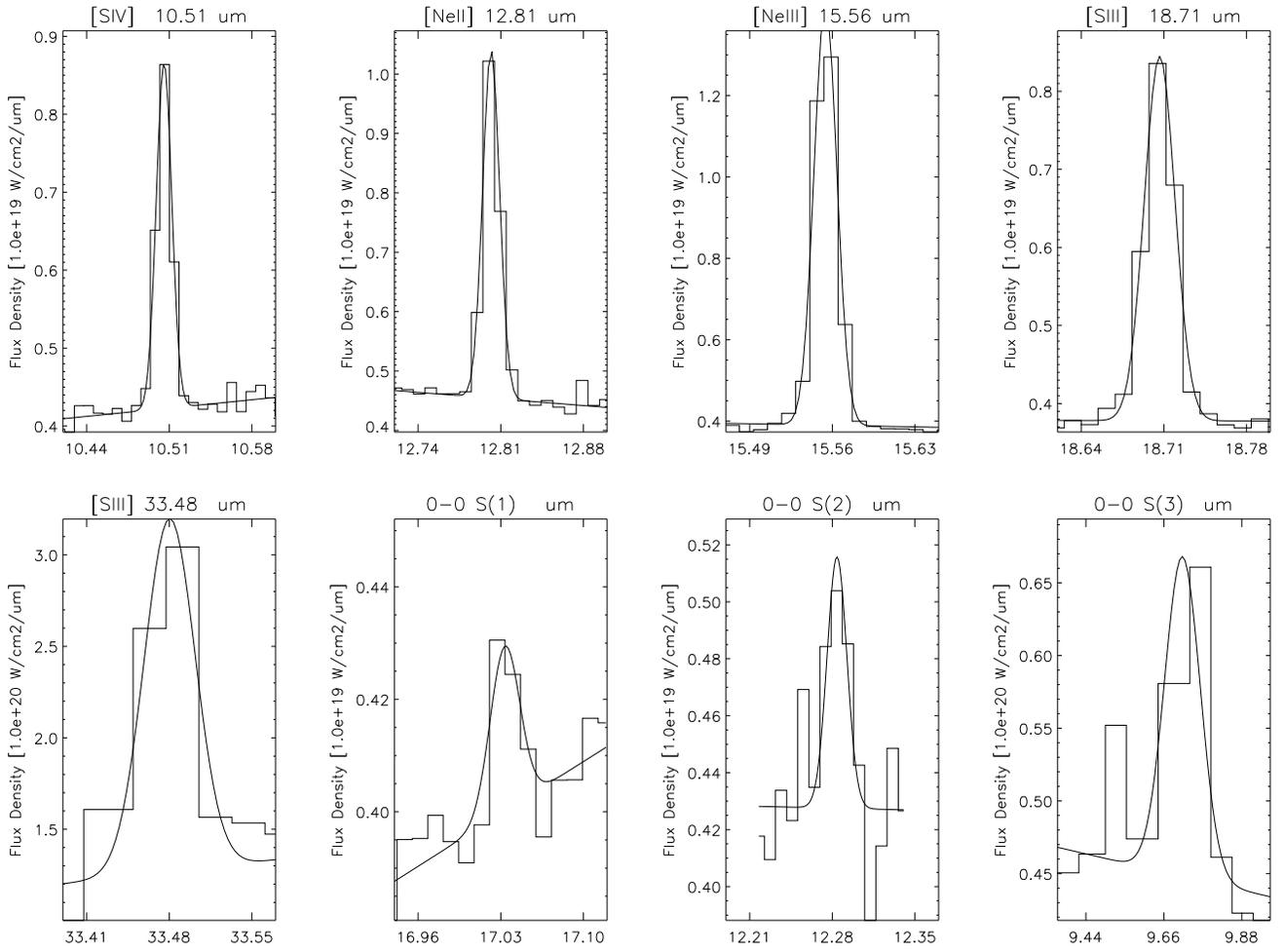}
\caption{Spitzer spectra of emission
  lines detected in TDG-N (histogram). The lines were fit with a
  Gaussian profile combined with a first order polynomial fit to the
  continuum (solid line). The lines are from IRS-SH/LH, except 0-0 S(3), 
which was observed with IRS-SL. }
\end{figure}

\clearpage


\begin{figure}
\centering
\label{fig:graphics:b}
\includegraphics[angle=90,scale=.75]{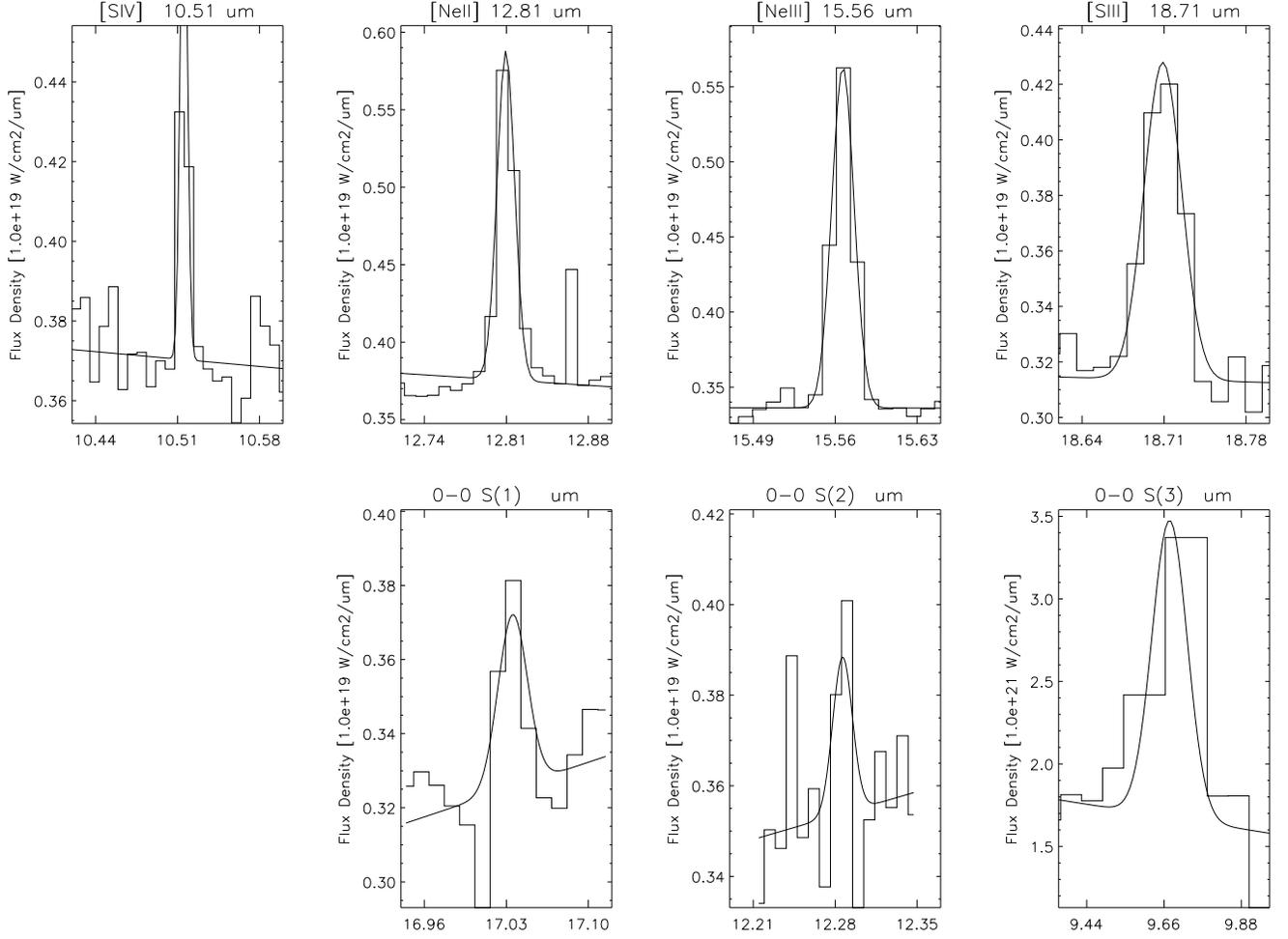}
\caption{ Spitzer IRS spectra of emission lines detected in TDG-S
  (histogram). The lines were fit with a Gaussian profile combined
  with a first order polynomial fit to the continuum (solid line).
  The lines are from IRS-SH/LH, except 0-0 S(3), which was observed
  with IRS-SL. All lines were detected in both nod positions except
  for the 0-0 S (2) line.}
\end{figure}

\clearpage

\begin{figure}
\centering
\includegraphics[angle=90,scale=.3]{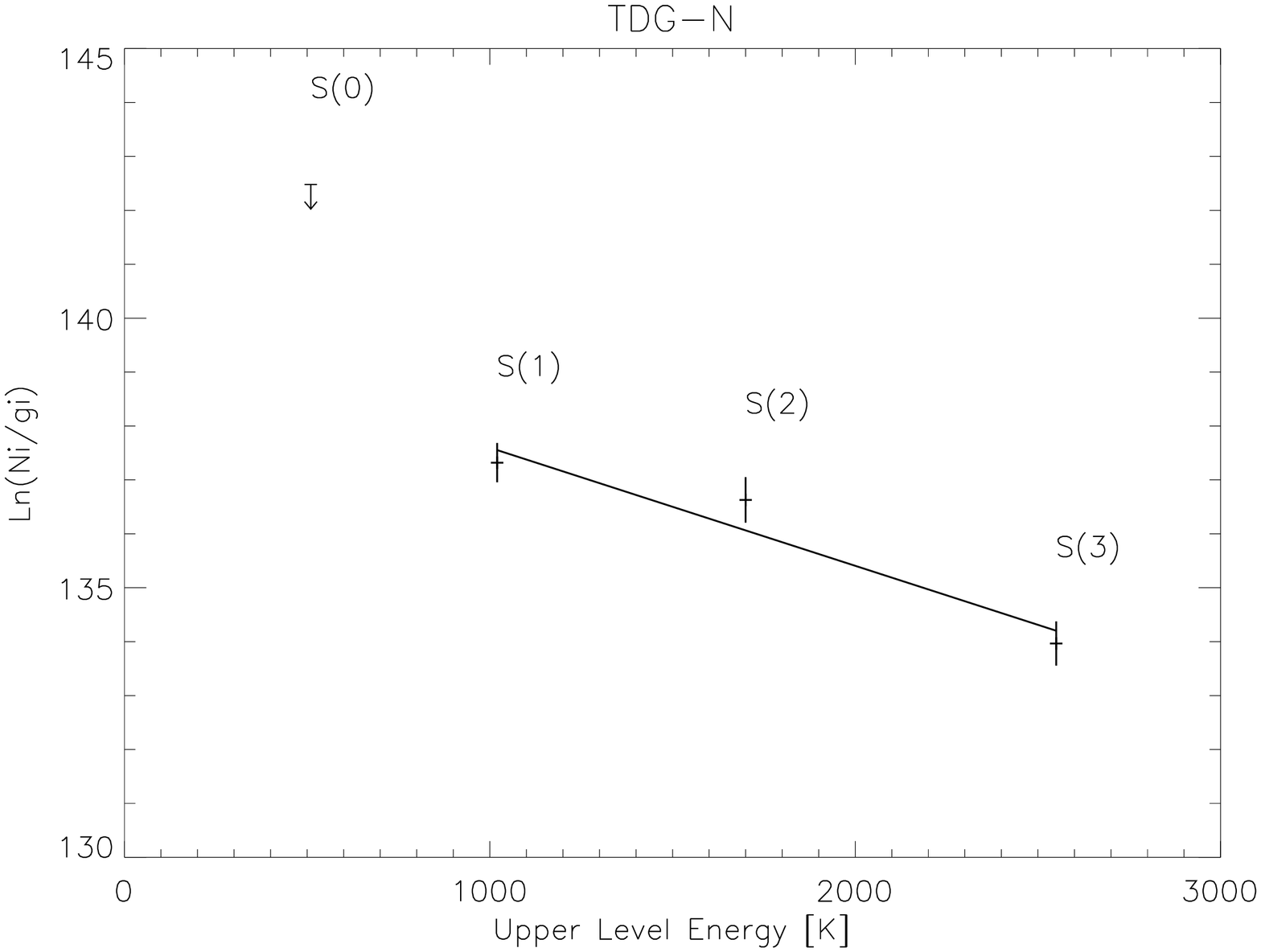}
\includegraphics[angle=90,scale=0.3]{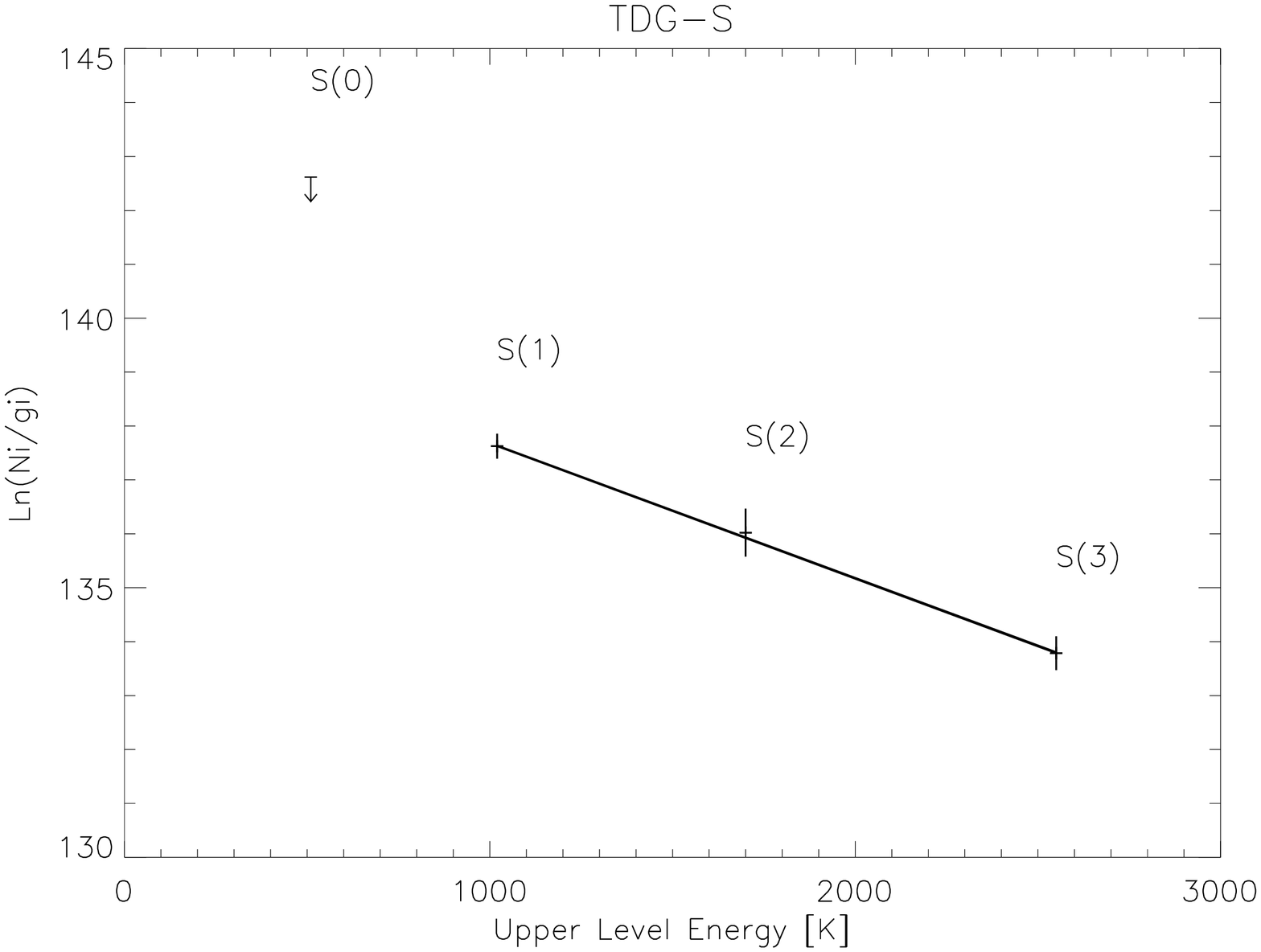}
\caption{Excitation Diagrams for TDG-N (left) and TDG-S (right). The
  absolute value of the reciprocal of the slope is the excitation
  temperature. }
\end{figure}

\clearpage

\begin{figure}
\centering
\caption{IRAC false color image of the NGC~5291 system, defined by 3.6
  \um (blue), 4.5 \um (green), 5.8 \um (orange) and 8.0 \um (red).
  Sources dominated by starlight (e.g, NGC 5291 \& `Seashell') are
  blue in color while objects with strong PAH emission, including TDGs
  and the flocculent spiral galaxy (\#5 in Table 4), are red in color.
}
\end{figure}

\clearpage

\begin{figure}
\centering
\epsscale{2.0}
\caption{(Top-left), IRAC 3.6 $\mu$m image of the NGC~5291 system. A
  linear stretch between -0.04 and 1.6 MJy sr$^{-1}$ is used. The
  scalebar represents 1\arcmin, i.e., a linear scale of 18 kpc.(Top-right),
  IRAC 4.5 $\mu$m image. A linear stretch between -0.05 and 1.2 MJy
  sr$^{-1}$ is used. (Bottom-left), IRAC 5.8 $\mu$m image. A linear stretch
  between -0.13 and 0.71 MJy sr$^{-1}$ is used. (Bottom-right), IRAC 8.0 $\mu$m
  image. A linear stretch between -0.10 and 0.72 MJy sr$^{-1}$ is
  used. }
\end{figure}

\clearpage
\epsscale{2.0}
\centerline{Fig. 9 --- cont.}

\clearpage
\epsscale{2.0}
\centerline{Fig. 9 --- cont.}

\clearpage

\epsscale{2.0}
\centerline{Fig. 9 --- cont.}

\clearpage

\begin{figure}
\centering
\caption{Close-ups of TDG-N (\# 26), TDG-S (\# 33) and candidate TDGs
  associated with NGC~5291 showing emission at 3.6 $\mu$m (top-row)
  and ``non-stellar'' emission from PAHs (bottom row). The vertical
  bar represents 30\arcsec, i.e., a linear scale of 9 kpc. }
\end{figure}
\clearpage

\begin{figure}
\centering
\label{fig:graphics:a}
\caption{IRAC images of NGC~5291 and the ``Seashell'' galaxy, showing
  (top-left) 3.6 $\mu$m emission, (top-right) 3.6 $\mu$m emission with
  surface brightness contours, (bottom-left) ``non-stellar'' emission,
  and (bottom-right) ``non-stellar'' emission after smoothing with a
  5\arcsec~ Gaussian kernel. The images are displayed with a
  logarithmic stretch between 0.1 and 1.0 MJy sr$^{-1}$, and the
  contours are equally spaced in log-space between 0.3 and 30.0 MJy
  sr$^{-1}$.}
\label{fig:graphics}
\end{figure}

\clearpage

\clearpage


\begin{figure}
\centering
\includegraphics[angle=90,scale=.75]{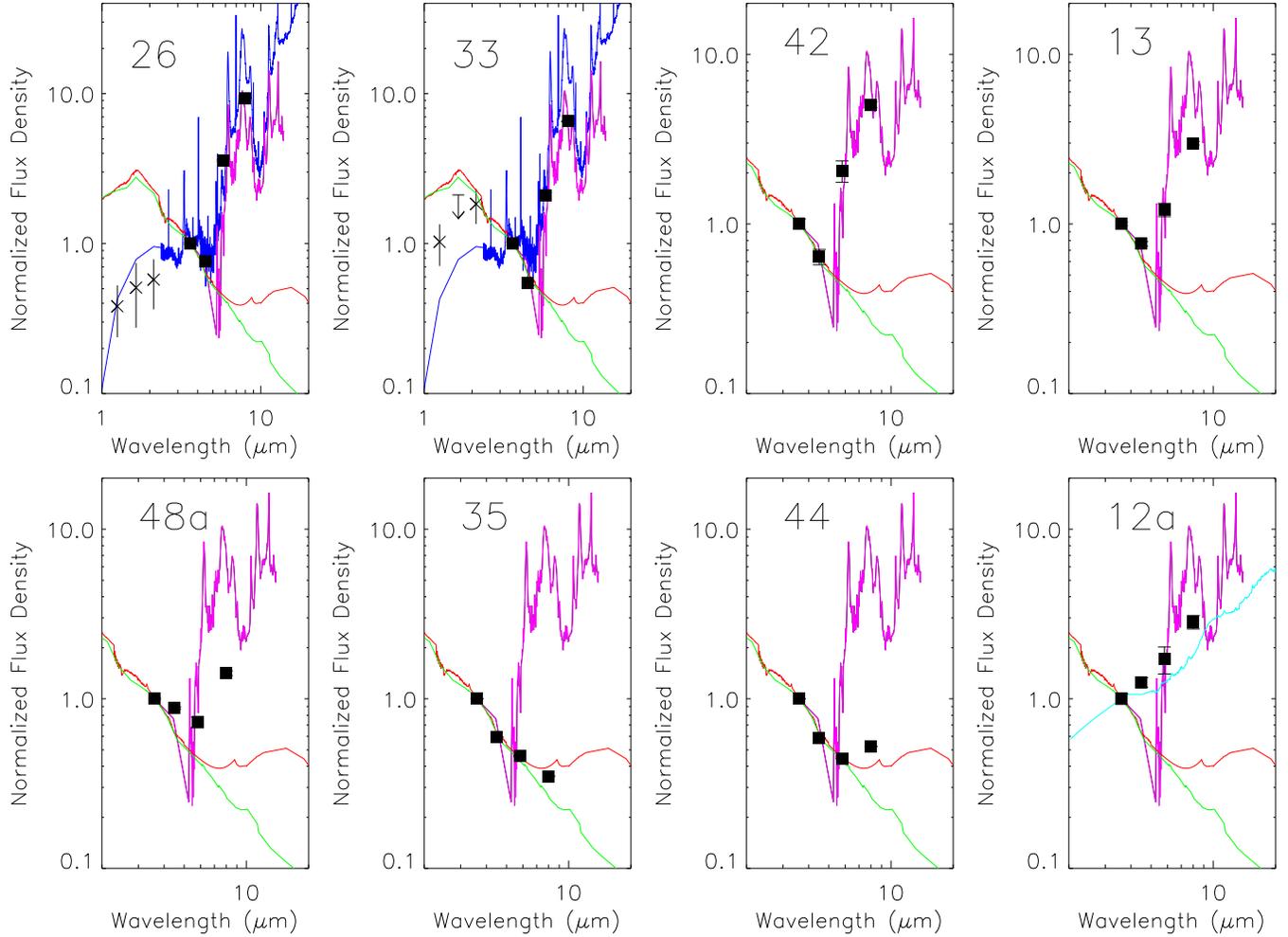}
\caption{Example IRAC Spectral Energy Distributions (SEDs) overlaid
  with a series of template galaxies: TDG-N the solid line (colored
  pink in electronic edition), Elliptical the dot-dash line (Silva et
  al.  1998, colored green in electronic edition), S0 the
  dot-dot-dot-dash line (Devriendt, Guiderdoni \& Sadat, R.  1999,
  colored red in electronic edition), M~82 nuclear region the dotted
  line (Lutz et al 1998, colored dark blue in electronic edition) and
  an I~Zw~1 the dash-line (Weedman et al. 2005, colored light blue in
  electronic edition).  Sources with the suffix `a' are outside the
  projected HI ring. TDG-N is \#26, TDG-S is \#33. \#42 is an example
  of a good match to TDG-N i.e., a candidate TDG, \#13 is a reasonable
  match to TDG-N and characterized as a star forming region, \#48a is
  a ``notched'' SED indicating star formation, the Seashell (\#35)
  matches both the elliptical and S0 templates, NGC~5291 (\#44) matches the
  S0 template and \#12a is a power-law, characteristic of an AGN.
}
\end{figure}

\clearpage

\begin{figure}
\centering
\label{fig:graphics:a}
\includegraphics[angle=90,scale=.75]{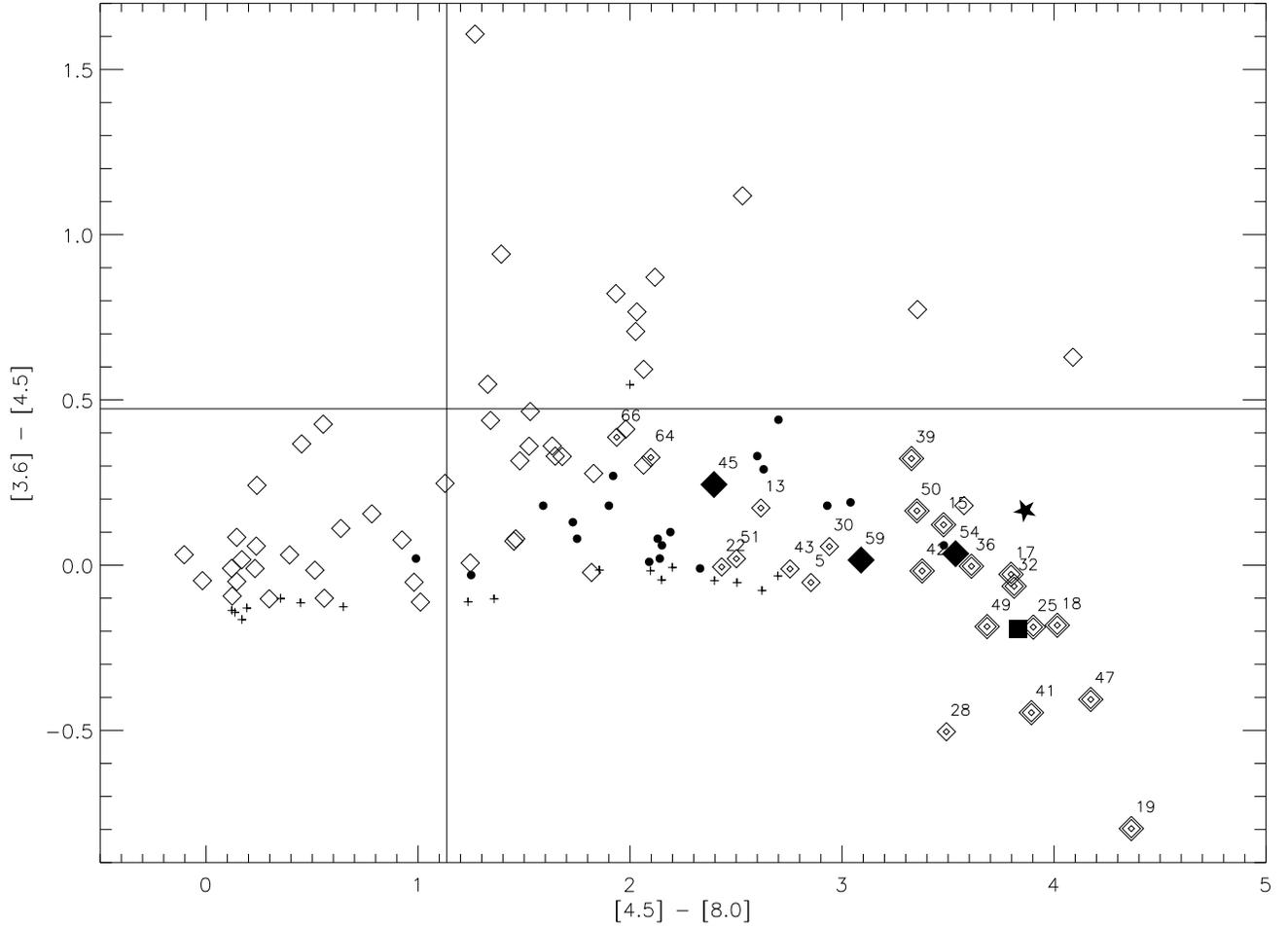}
\caption{ \iracb -\iracd versus \iraca - \iracb color $-$ color plot.
  The horizontal and vertical lines mark the color zero-points. A
  source with a flat SED would lie on the crossing point of these two
  lines. Sources in the lower left quadrant have falling SEDs typical
  of NGC~5291 and the Seashell, i.e., E and S0 galaxies. AGN like
  sources, with rising SEDs, are located in the upper right quadrant.
  The lower right quadrant contains sources with active star
  formation. TDG-N and TDG-S are shown as a small solid star and
  square, respectively. The diamonds represent the remaining sources
  from Table 4. In particular, framed-diamonds denote the 13 candidate
  TDGs. Sources located within the HI ring that are a reasonable match
  to TDG-N, but with weaker PAH emission, are shown as
  double-diamonds.  Sources outside the HI ring with TDG-like SEDs are
  shown as solid diamonds. Source identification numbers are from
  Table 4. For comparison the Pahre et al. (2004) galaxy sample are
  shown as crosses and the Rosenberg et al. (2005) BCDs are shown as
  solid circles. The TDGs and candidate TDGs all show significantly
  enhanced non-stellar emission, most likely due to PAHs, relative to
  normal spirals and even BCD galaxies. }
\end{figure}

\clearpage


\begin{figure}
\centering
\includegraphics[angle=90,scale=.75]{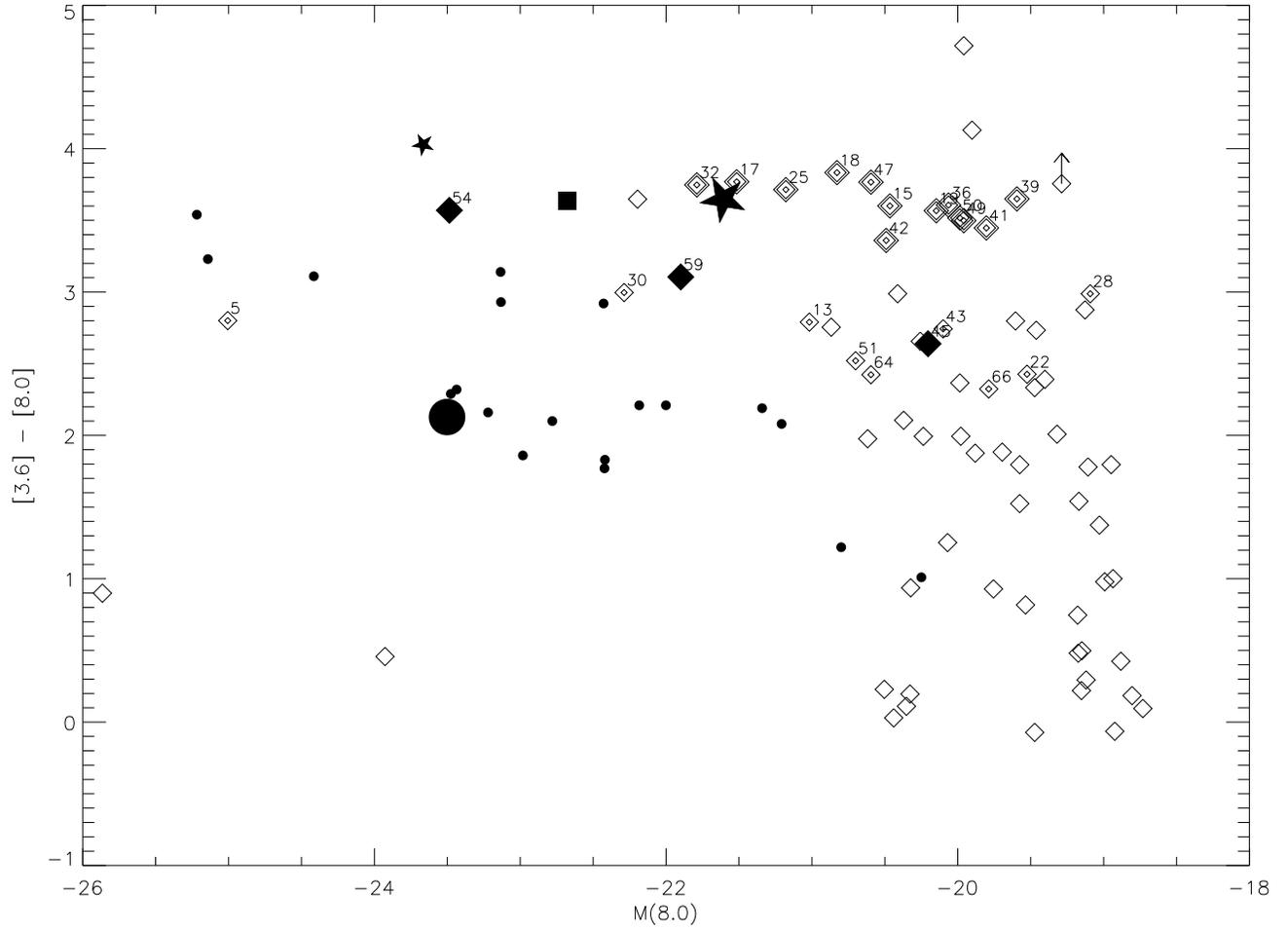}
\caption{\iraca - \iracd color versus \iracd absolute magnitude.  See
  the caption to Figure 13 for symbol definitions. The average
  positions of the TDG/TDG-candidate and BCD samples are indicated by
  the large star and circle, respectively. The candidate TDGs occupy a
  distinct region, being on average $\sim$ 2.2 magnitudes redder in
  \iraca $-$ \iracd color and 2.2 magnitudes fainter at 8 \um than
  the Rosenberg et al. (2005) BCD sample.}
\end{figure}

\clearpage

\end{document}